\documentclass{aa}

\usepackage{graphicx}

\usepackage{txfonts}
\usepackage{algorithm}
\usepackage{algpseudocode}
\usepackage{tikz}
\usetikzlibrary{bayesnet}
\usetikzlibrary{arrows}

\usepackage{xcolor}

\definecolor{mygreen}{HTML}{a3b899}
\definecolor{mygray}{HTML}{5d6d7e}
\definecolor{myblue}{HTML}{73c6b6}
\definecolor{mypink}{HTML}{cd6155}
\definecolor{myblue2}{HTML}{2980B9}

\newcommand{\sbilens}{\texttt{sbi\_lens}}

\usepackage[colorlinks=true,citecolor=myblue2,linkcolor=myblue2,urlcolor=myblue2]{hyperref}

\begin{document} 

\newcommand{\JZ}[1]{{\color{magenta} #1}}
\newcommand{\Ju}[1]{{\color{cyan}JZ: #1}}
\newcommand{\FL}[1]{{\color{magenta}FL: #1}}
\newcommand{\AB}[1]{{\color{olive}AB: #1}}
\renewcommand{\sectionautorefname}{Section}

   \title{
    Simulation-based inference benchmark for weak lensing cosmology}

    \author{
        Justine Zeghal \inst{1,9,10,11}
        \and
        Denise Lanzieri \inst{2,6}
        \and
        Fran\c{c}ois Lanusse \inst{3,7}
        \and
        Alexandre Boucaud \inst{1}
        \and 
        Gilles Louppe \inst{4}
        \and 
        Eric Aubourg \inst{5} 
        \and \\
        Adrian E.~Bayer \inst{8,7} 
        \and 
        The LSST Dark Energy Science Collaboration\
    }
    
    \institute{
        Université Paris Cité, CNRS, Astroparticule et Cosmologie, F-75013 Paris, France
        \and
        Université Paris Cité, Université Paris-Saclay, CEA, CNRS, AIM, F-91191, Gif-sur-Yvette, France
        \and
        Université Paris-Saclay, Université Paris Cité, CEA, CNRS, AIM, 91191, Gif-sur-Yvette, France
        \and 
        University of Liège, Liège, Belgium
        \and 
        Université Paris Cité, CNRS, CEA, Astroparticule et Cosmologie, F-75013 Paris, France
    \and 
 Sony Computer Science Laboratories - Rome, Joint Initiative CREF-SONY, Centro Ricerche Enrico Fermi, Via Panisperna 89/A, 00184, Rome, Italy
 \and Center for Computational Astrophysics, Flatiron Institute, 162 5th Ave, New York, NY, 10010, USA
 \and 
Department of Astrophysical Sciences, Princeton University, Peyton Hall, 4 Ivy Lane, Princeton, NJ 08544, USA
\and Department of Physics, Université de Montréal, Montréal H2V 0B3, Canada
 \and Mila – Quebec Artificial Intelligence Institute, Montréal H2S 3H1, Canada
 \and Ciela – Montreal Institute for Astrophysical Data Analysis and Machine Learning, Montréal H2V 0B3, Canada
}

   \date{Received XXXX; accepted XXXX}

  \abstract
   {Standard cosmological analysis, which is based on two-point statistics, fails to extract all the information embedded in the cosmological data.  This limits our ability to precisely constrain cosmological parameters.  Through willingness to use modern analysis techniques to match the power of upcoming telescopes, recent years have seen a paradigm shift from analytical likelihood-based to simulation-based inference. However, such methods require a large number of costly simulations.}
   {We focused on full-field inference, which is considered the optimal form of inference as it enables the recovery of cosmological constraints from simulations without any loss of cosmological information. Our objective is to review and benchmark several ways of conducting full-field inference to gain insight into the number of simulations required for each method. Specifically, we made a distinction between explicit inference methods that require an explicit form of the likelihood, such that it can be evaluated and thus sampled through sampling schemes and implicit inference methods that can be used when only an implicit version of the likelihood is available through simulations. 
   Moreover, it is crucial for explicit full-field inference to use a differentiable forward model. Similarly, we aim to discuss the advantages of having differentiable forward models for implicit full-field inference.}
   {We used the \href{https://github.com/DifferentiableUniverseInitiative/sbi_lens}{\sbilens} package, which provides a fast and differentiable log-normal forward model to generate convergence maps mimicking a simplified version of LSST Y10 quality. While the analyses use a simplified forward model, the goal is to illustrate key methodologies and their implications. Specifically, this fast-forward model enables us to compare explicit and implicit full-field inference with and without gradient. The former is achieved by sampling the forward model through the No U-Turns (NUTS) sampler. The latter starts by compressing the data into sufficient statistics and uses the neural likelihood estimation (NLE) algorithm and the one augmented with gradient ($\partial$NLE) to learn the likelihood distribution and then sample the posterior distribution.}
   {We performed a full-field analysis on LSST Y10-like weak-lensing-simulated log-normal convergence maps, where we constrain $(\Omega_c, \Omega_b, \sigma_8, h_0, n_s, w_0)$. We demonstrate that explicit full-field and implicit full-field inference yield consistent constraints. Explicit full-field inference requires $630\,000$ simulations with our particular sampler, which corresponds to $400$ independent samples. Implicit full-field inference requires a maximum of $101\,000$ simulations split into $100\,000$ simulations to build neural-based sufficient statistics (this number of simulations is not fine-tuned) and $1\,000$ simulations to perform inference using implicit inference. Additionally, while differentiability is very useful for explicit full-field inference, we show that, for this specific case, our way of exploiting the gradients does not help implicit full-field inference significantly.}
   {}


   \maketitle

\section{Introduction}

Understanding the cause of the observed accelerated expansion of the Universe is currently a major topic in cosmology. The source of this acceleration has been dubbed dark energy, but its nature is still unknown. Dark energy cannot be directly observed, but several observational probes can be used to better understand its characteristics; weak gravitational lensing, in which background galaxies are sheared by foreground matter, is among the most powerful. This phenomenon is sensitive to both the geometry of the Universe and the growth of structure, which both depend on the cosmological parameters of the dark-energy model. Many photometric galaxy surveys such as CFHTLenS \citep{cfht}, KiDS \citep{kids}, DES \citep{des}, and HSC \citep{hsc} have already demonstrated its constraining power on the matter-density $\Omega_m$ and fluctuation-amplitude $\sigma_8$ parameters. Upcoming weak lensing surveys (LSST \citep{lsst}, Roman \citep{roman}, Euclid \citep{euclid}) are expected to be larger and deeper, allowing us to refine our estimations even further. 

In cosmological inference, a significant challenge lies in the absence of an analytic likelihood $p(x| \theta)$ to recover cosmological parameters $\theta$ from the data $x$. Most of the mathematical inference frameworks proposed to overcome this problem are based on two-stage inference: compression of the data into summary statistics $t=f(x)$ and then Bayesian inference to obtain the posterior $p(\theta| t)$. The most famous is the two-point statistics analysis (e.g., \citealp{2pt_stat}). It uses the two-point correlation function or its analog in Fourier space, the power spectrum, as a summary statistic $t$.  Then, the inference part of the analysis is performed using the corresponding analytic Gaussian likelihood $p(t| \theta)$, which is sampled using a Markov chain Monte Carlo (MCMC) method. On large scales, the Universe remains close to a Gaussian field, and the two-point function is a near-sufficient statistic to extract cosmological information. However, on small scales where nonlinear evolution gives rise to a highly non-Gaussian field, this summary statistic is no longer sufficient. 

At a time when future surveys will access small scales, we need to investigate summary statistics that can capture non-Gaussianities. This has led to a new class of statistics, known as higher order statistics, including, for example,  lensing peak counts (e.g., \citealp{liu2015cosmology,  liu2015cosmological, lin2015new, kacprzak2016cosmology, peel2017cosmological, shan2018kids, martinet2018kids, ajani2020constraining, harnois2021cosmic, zurcher2022dark}), three-point statistics (e.g., \citealp{takada2004cosmological, semboloni2011weak, Fu_2014, rizzato2019tomographic, halder2021integrated}), and machine-learning compression (e.g., \citealp{Charnock_2018, PhysRevD.98.123518, 2018PhRvD..97j3515G,  ribli2019weak, jeffrey2021likelihood, fluri2022full, Akhmetzhanova2024, Jeffrey2024}), all with varying degrees of signal-extraction power. Most of the time, no analytical models $p(t|x)$ exist, and these statistics are usually assumed to be Gaussian-distributed, leading to potentially biased inference or an inaccurate uncertainty estimation. On top of that, since no analytical function $t = g(\theta)$ to map cosmological parameters to the summary statistic exists, the inference part requires a large number of very costly simulations $x \sim p(x|\theta)$ (with $p(x|\theta)$ a simulator) to compute the summary statistics $t = f(x)$. This is in addition to the number of simulations already required to compute the covariance matrix. 

Full-field inference (e.g., \citealp{schneider2015hierarchical, alsing2016hierarchical, alsing2017cosmological, bohm2017bayesian, porqueres2021bayesian, porqueres2022lifting, porqueres2023field, Zhou2023, dai2024multiscale, lanzieri2024optimalneuralsummarisationfullfield}) aims to perform inference from simulations without any loss of information. This means no loss of information coming from a compression step and no loss of information coming from assumptions on the likelihood function employed for inference. Hence, the quality of the learned posterior is solely tied to the forward model's accuracy. This paper focuses on this particular kind of inference.

\vspace{2mm}

Depending on the nature of the forward model $p(x|\theta)$, one can either perform explicit inference or implicit inference. The former refers to inference methods that can be used when the likelihood function $p(x = x_0|\theta)$ can be evaluated for different $\theta$, and sampling schemes can thus be employed. The latter can be used when only an implicit version of the likelihood is available through a set of simulations $(\theta, x)$. Implicit inference, also known as likelihood-free inference or simulation-based inference, commonly recasts the inference problem as a neural-density optimization problem where the distribution is learned and can be evaluated for all $\theta$ and $x$. Different types of implicit inference exist; one aims to learn the likelihood function $p(x |\theta)$ (e.g., \citealp{wood_statistical_2010, nle1, nle2, SNPSE}) or the likelihood ratio $r(\theta, x) = p(x |\theta) \: / \: p(x)$ (e.g., \citealp{lr2,lr1,lr3,hermans2020likelihoodfree,lr4, miller2023contrastive}). This learned likelihood or likelihood ratio can now be evaluated, and sampling methods can be used to obtain the posterior $p(\theta | x)$. Others choose to directly approximate the posterior distribution $p(\theta | x)$ (e.g., \citealp{pe,npe1,npe2,npe3, wildberger2023flow}).

Explicit inference applied in the context of full-field inference is known as Bayesian hierarchical/forward modeling. Because of the high complexity and dimension of the field-based likelihood, sampling schemes guided by the gradient information $\nabla_{\theta} \log p(x| \theta)$ are typically used to explore the parameter space in a more efficient way. This motivates the development of differentiable forward models $p(x|\theta)$.
Naturally, we could ask whether these gradients also help implicit inference methods for full-field inference. Specifically, \cite{mining_gold} and \cite{zeghal2022} proposed implicit inference methods to leverage the gradient information from the forward model while approximating the likelihood, the likelihood ratio, or the posterior distribution. They showed that this additional information helps constrain the target distribution and thus improve sample efficiency. 

\vspace{2mm}

In summary, this paper aims to determine if the differentiability of the forward model is a useful asset for full-field implicit inference and which methods allow full-field inference with the fewest simulations. To meet the full-field criterion, we focused our benchmark analysis on two inference strategies. The first is explicit full-field inference, whereby we sampled our forward model through the Hamiltonian Monte Carlo (HMC) sampling method. Specifically, we use the No-U-Turn (NUTS) algorithm. The second is implicit full-field inference; after compressing the simulations into sufficient statistics, we compared the neural likelihood estimation (NLE) and neural likelihood estimation augmented with gradients ($\partial$NLE).

For the implicit inference strategy, maps were compressed using an optimal neural compression approach: we trained a convolutional neural network (CNN) by maximizing the mutual information between the cosmological parameters and the summary statistic (e.g., \citealp{jeffrey2021likelihood}; see \citealp{lanzieri2024optimalneuralsummarisationfullfield} for a review on optimal neural compression strategies). In this study, we separated the compression process from the inference process and concentrated solely on the amount of simulations necessary for inference. We explain why in Sect. \ref{sec:compression}.

We used the same forward model to benchmark the different inference strategies and use the same fiducial data $x_0$. Our forward model is a differentiable field-based likelihood that can be evaluated and can generate simulations such that both approaches, explicit and implicit, can be performed. Specifically, it is a log-normal model that produces LSST Y10-like weak lensing convergence maps. The cosmological parameters $\theta$ that we aim to constrain are $(\Omega_c, \Omega_b, \sigma_8, h_0, n_s, w_0)$. The forward model can be found in \href{https://github.com/DifferentiableUniverseInitiative/sbi_lens}{\sbilens}.\\
\begin{figure*}[h]
        \begin{center}
        \includegraphics[width=\textwidth]{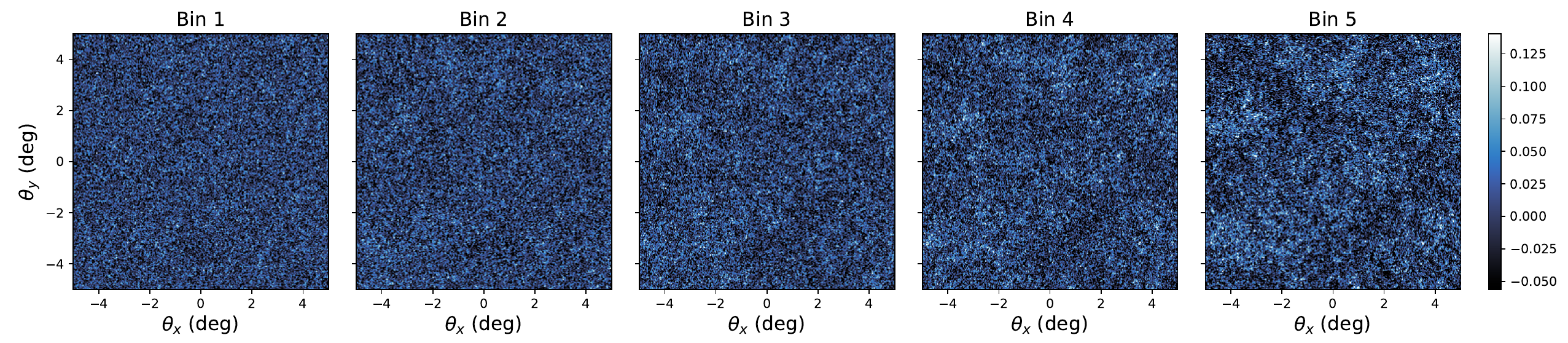}
        \caption{Example of noisy convergence map, with five tomographic redshift bins, generating with \sbilens's log-normal forward model. It is the fiducial map $x_0$ used for benchmarking all the inference techniques.}
        \label{fig:ex_kmap}
         \end{center}
    \end{figure*}

We start by introducing our lensing forward modeling in Sect. \ref{sec:forwardmodel}. In Sect. \ref{sec:bayesian_inference}, we introduce our Bayesian inference framework. Then in Sect. \ref{sec:metric}, we present the metric used to benchmark the different inference approaches. We then describe the explicit inference approach and present the results in Sect. \ref{sec:bhm}. It is followed by the implicit inference approaches both with and without gradients and the corresponding results in Sect. \ref{sec:implicit_inference}. Finally, we conclude in Sect. \ref{sec:conclusion}.

\section{The lensing forward model}
    \label{sec:forwardmodel}

    Due to the nonlinear growth of structures in the Universe, the cosmological density field is expected to be highly non-Gaussian. Therefore, log-normal fields that account for non-Guassianities\footnote{Figure \ref{figappendix:lognormal} quantifies the amount of non-Gaussianities in our model compared to Gaussian simulations.} provide a fast representation of the late-time 2D convergence field \citep{xavier2016improving, clerkin2017testing}. 
    
    For our study, we used \href{https://github.com/DifferentiableUniverseInitiative/sbi_lens}{\sbilens}’s JAX-based differentiable forward model introduced in \cite{lanzieri2024optimalneuralsummarisationfullfield} to generate log-normal simplified LSST Y10 convergence maps. In this section, we recall the log-normal forward model of \cite{lanzieri2024optimalneuralsummarisationfullfield}.

    \subsection{Log-normal modeling}
    
    Given a Gaussian field $\kappa_g$ fully characterized by its correlation function $\xi^{ij}_{g}$ ($i$ and $j$ denoting the $i$-th and $j$-th source redshift bins see Sect. \ref{sec:lssty10}), we parametrize the log-normal field $\kappa_{ln}$ as 
    \begin{equation} \label{eq:lognormalfield}
        \kappa_{ln}=e^{\kappa_{g}}-\lambda,
    \end{equation}
    with $\lambda$ being an additional parameter that makes the log-normal field more flexible than its corresponding Gaussian field. This parameter is called the "shift" or “minimum value" and depends on the cosmology.
    Hence, the field is no longer only described by its correlation function.

    We note that this log-normal transformation leads to the following modification of the correlation function: 
    \begin{equation}
        \xi^{ij}_{ln} = \lambda_i \lambda_j (e^{ \xi^{ij}_g}-1).
    \end{equation}
To ensure that the log-normal field shares the same correlation function as its Gaussian analog, we applied the following correction:
    \begin{equation}\label{eq:correction}
        f(\xi^{ij}) = \log{\left[ \frac{\xi^{ij}}{\lambda_i \lambda_j}+1\right ]},
    \end{equation}
which also makes the correlation function independent of the choice of the shift parameter. However, the shift parameter has to be carefully set as it is related to the skewness of $\kappa_{ln}$. It can be computed from simulations using matching moments \citep{Xavier_2016} or by using perturbation theory \citep{cosmomentum}.
    
Finally, the correlation function is related to the power spectrum by
    \begin{equation}\label{Eq:log_norm_cls}
        C^{ij}_{ln}(\ell)=2\pi \int_0^{\pi} d\theta \sin{\theta}P_{\ell}(\cos{\theta})\xi^{ij}_{ln}(\theta),
    \end{equation}
with $P_{\ell}$ being the Legendre polynomial of order $\ell$. In Fourier space, the covariance of $\kappa_{ln}$ is diagonal and defined as
\begin{equation}\label{power_spectrum_definition}
    \left \langle \tilde{\kappa}^{(i)}_{ln} (\ell) \: \tilde{\kappa}^{*(j)}_{ln}(\ell')\right \rangle =C^{ij}_{ln}(\ell)\delta^{K}(\ell-\ell').
    \end{equation}
    \subsection{\sbilens's log-normal forward model}
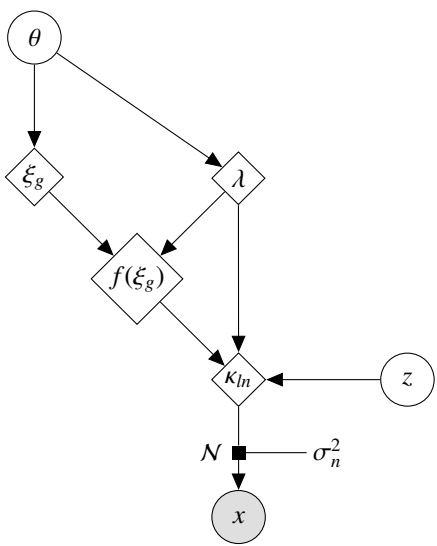
\begin{figure}[h]
        \centering
        \begin{tikzpicture}
          \node[obs]          (y)   {$x$}; %
          \factor[above=of y] {y-f} {left:$\mathcal{N}$} {} {} ; %
        
          \node[const, right=0.8 of y-f]       (noise)   {$\: \sigma_n^2$};
          \node[det, above=1.1 of y]         (k)   {$\kappa_{ln}$};
          \node[det, above left=1.2 of k]    (dot) {$f(\xi_g)$} ; %
          \node[det, above left=1.2 of dot]  (w)   {$\xi_{g}$}; %
          \node[det, above right=1.2 of dot] (x)   {$\lambda$}; %
          \node[latent, above=1.1 of w]  (theta)   {$\theta$}; %
        
          \node[latent, right=1.5cm of k]         (t)   {$z$}; %
        
          \factoredge {k,noise} {y-f} {y} ; %
        
          \edge {theta} {w,x} ;
          \edge {w,x} {dot} ;
          \edge {dot, x, t} {k} ;
        
        \end{tikzpicture}
        \caption{Representation of \sbilens's forward model used to generate log-normal convergence maps.}
        \label{fig:bhm}
    \end{figure}

    \sbilens's forward model is structured as follows (see  Fig. \ref{fig:bhm}): first,  we define the prior $p(\theta)$ over the cosmological parameters $(\Omega_c, \Omega_b, \sigma_8, n_s, w_0, h_0)$ (see \autoref{tab:fid&prior}). Given a cosmology from the prior, we compute the corresponding nonlinear power spectrum $C_{\ell, g}$ using \href{https://github.com/DifferentiableUniverseInitiative/jax_cosmo}{\url{JAX-COSMO}} \citep{Campagne_2023}, which we project on two-dimensional grids of the size of the final mass map. For this cosmology, we also compute the cosmology-dependent shift parameter $\lambda$ using 
\href{https://github.com/OliverFHD/CosMomentum}{\url{CosMomentum}} \citep{cosmomentum}. To ensure that the log-normal field preserves the power spectrum $C_{\ell, g}$, we applied the correction to the correlation function from Eq. \ref{eq:correction}.
Then, we convolve the Gaussian latent variables
 $z$ (also known as latent variables) with the corrected two-dimensional power spectrum:
 \begin{equation}
     \hat{\kappa_g} = \hat{Z} \cdot \Sigma^{1/2},
 \end{equation}
with $\hat{Z}$ denoting the Fourier transform of the latent variables $z$ and $\Sigma^{1/2}$ the square root of the covariance matrix
  \begin{equation}
    \Sigma=  (C_{\ell}^{ij})_{1\leq i\leq 4, 1 \leq j \leq 4}.
\end{equation} 
Here, $C_{\ell}^{ij}$ denotes the corrected and projected nonlinear power spectrum. To compute the square root of the covariance matrix, we performed an eigenvalue decomposition of $\Sigma$: 
\begin{equation}
\Sigma=Q \Lambda Q^{T},
\end{equation}
with $Q$ being the eigenvectors and $\Lambda$ the eigenvalues of the symmetric matrix $\Sigma$. This allowed us to compute the square root efficiently as
\begin{equation}
\Sigma^{1/2}=Q\Lambda^{1/2}Q^{T}.
\end{equation}
 Finally, we built the log-normal field $\kappa_{ln}$ as described by  Eq. \ref{eq:lognormalfield}. An example of log-normal convergence maps is shown in Fig. \ref{fig:ex_kmap}.
\begin{table}
 \caption{Prior and fiducial values used for our inference benchmark.}
        \begin{center}
        \begin{tabular}{|c|c|c|} 
                \hline 
                Parameter  & Prior & Fiducial value \\
            \hline \hline
                $\Omega_c$ & $\mathcal{N}_{T[0, +\infty]} (0.2664, 0.2)$ & 0.2664 \\
                $\Omega_b$ & $\mathcal{N} (0.0492, 0.006)$ & 0.0492 \\
                $\sigma_8$ & $\mathcal{N} (0.831, 0.14)$ & 0.8310 \\
                $h_0$ & $\mathcal{N} (0.6727, 0.063)$ & 0.6727\\
                $n_s$ & $\mathcal{N} (0.9645, 0.08)$ & 0.9645 \\
                $w_{0}$ &  $\mathcal{N}_{T[-2.0, -0.3]} (-1.0, 0.9)$ &  -1.0 \\
                \hline
        \end{tabular}
        \tablefoot{Prior $p(\theta)$ used in our forward model and fiducial values used for our inference benchmark. $\mathcal{N}_{T[low, \: high]}$ refers to truncated normal distributions between $low$ and $high$. The priors and fiducial values are the same as LSST DESC SRD.}
            \label{tab:fid&prior}
    \end{center}
\end{table}

\subsection{LSST Y10 settings}
\label{sec:lssty10}
According to the central limit theorem, we assume LSST Y10 observational noise to be Gaussian as we expect a high number of galaxies per pixel. Hence, the shear noise per pixel is given by a zero-mean Gaussian whose standard deviation is
\begin{equation}
    \sigma^2_n= \frac{\sigma_e^2}{N_s}, 
    \label{eq:noise}
\end{equation}
where $\sigma_e = 0.26$ is the per-component-shape standard deviation as defined in the LSST DESC Science Requirement Document (SRD, \citet{mandelbaum2018lsst}), and $N_s$ is the number of source galaxies per bin and pixel, computed using $n_{gal}=27$ arcmin$^{-2}$ the galaxy number density (as in LSST DESC SRD) and $A_{pix}\approx 5.49$ arcmin$^2$ the pixel area. 
The convergence field is related to the shear field through the Kaiser Squires operator \citep{kaiser1993mapping}. As this operator is unitary, it preserves the noise of the shear field; therefore, the convergence noise is also given by Eq. \ref{eq:noise}.

\begin{figure}[h]
    \includegraphics[width=9.cm]{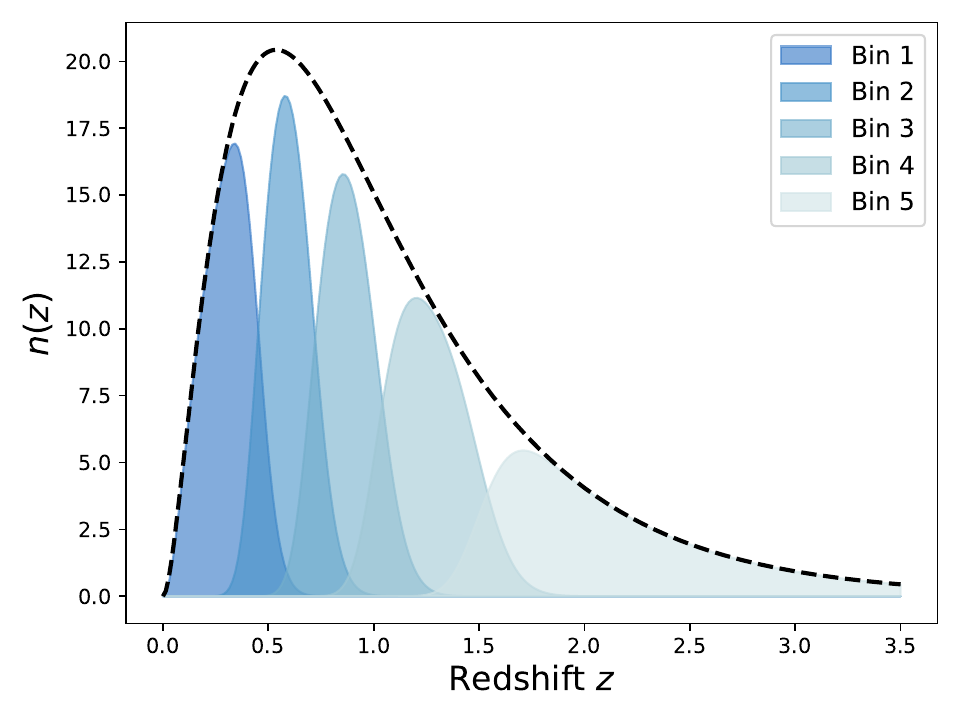}
    \caption{LSST Y10 redshift distribution used in our forward model.}
    \label{fig:redshift}
\end{figure}

 Our convergence map, $x$, is a $256 \times 256$ pixel map that covers an area of $10 \times 10$ deg$^2$ in five tomographic redshift bins with an equal number of galaxies (see Fig. \ref{fig:redshift}). The redshift distribution is modeled using the parametrized Smail distribution \citep{smail1995deep}:
 \begin{equation}
    n(z)\propto z^2 \exp{-(z/z_0)^{\alpha}},
\end{equation}
with $z_0=0.11$ and $\alpha=0.68$. We assume a photometric redshift error of $\sigma_z=0.05(1+z)$ (still according to LSST DESC SRD).

\section{Bayesian inference}
\label{sec:bayesian_inference}

    In this section, we introduce our Bayesian inference framework enabling us to distinguish between implicit and explicit (full-field) inference more clearly in this paper.
    
    Given a priori knowledge $p(\theta)$ about the parameters $\theta$ and information provided by data $x$ linked to the parameters via the likelihood function $p(x | \theta)$, we are able to recover the parameters $\theta$ that might have led to this data. This is summarized by Bayes' theorem: 
    \begin{equation}
    \label{eq:bayes}
        p(\theta |x) = \frac{p(x |\theta)p(\theta)}{p(x)}, 
    \end{equation}
with $p(\theta|x)$ being the posterior distribution of interest and $p(x) = \int p(x|\theta)p(\theta)d\theta$  the evidence. However, physical forward models are typically of the form $p(x| \theta, z)$, involving additional variables $z$ known as latent variables. The presence of these latent variables makes the link between the data $x$ and the parameters $\theta$ not straightforward, as $x$ is now the result of a transformation involving the two random variables $\theta$ and $z$. Since the forward model depends on latent variables, we need to compute the marginal likelihood to perform inference; that is,
    \begin{equation}
        p(x  | \theta) = \int p(x |\theta, z)p(z | \theta)dz,
        \label{eq:marginal_likelihood}
    \end{equation}
which is typically intractable when $z$ is of high dimension. As a result, the marginal likelihood $p(x  | \theta)$ cannot be evaluated, and explicit inference techniques that rely on explicit likelihood such as the MCMC method or variational inference cannot be directly applied to the marginal likelihood $p(x  | \theta)$. For this reason, this marginal likelihood is often assumed to be Gaussian, yielding an inaccurate estimation of the true posterior. Full-field inference instead aims to consider the exact distribution of the data $x$ or the sufficient statistics $t$. 

\section{Inference quality evaluation}
\label{sec:metric}
    To quantify the quality of inference and thus benchmark all the inference algorithms, a performance metric has to be carefully chosen. Several metrics exist, each offering varying levels of precision, and they are usually chosen according to the knowledge we have about the true posterior (i.e., if we have access to the probability density function of the true distributions, its samples, or only the fiducial data or fiducial parameters).
    We took the $160\,000$ posterior samples obtained through explicit full-field inference as our ground truth and used the Classifier 2-Sample Tests (C2ST, \citealp{lopezpaz2018revisiting}). This decision is based on the understanding that the explicit full-field approach, which relies on sampling schemes, should theoretically converge to the true posterior distribution within the limit of a large number of samples. 
    The convergence analysis of the MCMC, along with the large number of samples ($160\,000$), indicates that the explicit, full-field inference posterior has converged. Additionally, we confirm this by visually comparing (in Fig. \ref{fig:conv_contour}) the marginals of the fully converged samples obtained through explicit full-field inference (black) to the marginals of the posterior obtained through implicit inference (blue).  Although we present implicit inference performed with only $1\,000$ simulations in Fig. \ref{fig:conv_contour}, it is worth noting that we ran the implicit inference method with over $1\,000$ simulations, and it was consistently in agreement with this explicit posterior.

    We selected the C2ST metric due to its recognition as an effective and interpretable metric for comparing two distributions, particularly in the context of implicit inference. According to \cite{lueckmann2021benchmarking}, C2ST outperforms other metrics such as the maximum mean discrepancy (MMD) and posterior predictive checks (PPCs) in various benchmarks. For instance, in the "Two Moons" benchmark task discussed by \cite{lueckmann2021benchmarking}, C2ST demonstrated superior sensitivity to differences in distributions compared to MMD, which was found to be overly sensitive to hyperparameter choices. Furthermore, C2ST proved effective in scenarios involving multi-modal and complex posterior structures, where other metrics struggled to provide consistent and reliable results.

    A two-sample test is a statistical method that tests whether samples $X \sim P$ and $Y \sim Q$ are sampled from the same distribution. For this, one can train a binary classifier $f$ to discriminate between $X$ (label 0) and $Y$ (label 1) and then compute the C2ST statistic 
    \begin{equation}
        \hat{t} = \frac{1}{N_{test}} \sum_{i=1}^{N_{test}} \mathbb{I} \left[\mathbb{I} \left(f(z_i) > \frac{1}{2}\right) = l_i \right], 
    \end{equation}
where $\{(x_i, 0)\}_{i=1}^N \cup \{(y_i, 1)\}_{i=1}^N =: \{(z_i, l_i)\}_{i=1}^{2N}$ and ${N_{test}}$ denotes the number of samples not used during the classifier training. If $P = Q$, the classifier fails to distinguish the two samples and thus the C2ST statistic remains at chance level (C2ST = 0.5). On the other hand, if $P$ and $Q$ are so different that the classifier perfectly matches the right label, C2ST = 1.  
    
    In practice, in our six-dimensional inference problem, we find this metric very sensitive, and the two distributions considered converged in Fig. \ref{fig:conv_contour} result in a C2ST of $0.6$. Therefore, to make a fair comparison between all inference methods we benchmark all the methods with the same metric, the C2ST metric, and choose to fix a threshold of $0.6$.

\section{Explicit inference}
\label{sec:bhm}
\subsection{Sampling the forward model}
    When the forward model is explicit, which means that the joint likelihood $p(x | \theta, z)$
    can be evaluated, it is possible to sample it directly through an MCMC method, bypassing the computation of the intractable marginal likelihood $p(x | \theta)$. 

    Unlike sampling the marginal likelihood, this necessitates sampling both the parameters of interest $\theta$ as well as all latent variables $z$ involved in the forward model, 
    \begin{equation}
        p(\theta, z | x) \propto  p(x | \theta, z) \: p(z | \theta) p(\theta),
    \end{equation}
    and marginalizing over the latent variables $z$ afterward to obtain the posterior distribution $p(\theta |x)$. 
    
    As the latent variables are usually high-dimensional, they require a large number of sampling steps to make the MCMC converge. Therefore, Hamilton Monte Carlo (HMC, \citealp{neal2011mcmc, betancourt2018conceptual}), which can efficiently explore the parameter space thanks to gradient information, is usually used for such high-dimensional posteriors. However, this requires the explicit likelihood to be differentiable. 
    
    We note that for each step the forward model needs to be called, which can make this approach costly in practice as generating one simulation can take a very long time. This would also be true in cases where the marginal likelihood can be evaluated, but since the latent variables $z$ do not have to be sampled, the parameter space is smaller and the MCMC does not need as many steps.

    \subsection{Explicit full-field inference constraints}
    \sbilens' differentiable joint likelihood is 
    \begin{equation}
        p(x | \theta, z) = \mathcal{N}\left(\kappa_{ln}(\theta, z), \sigma^2_n\right),
    \end{equation}
    with $\kappa_{ln}$ being the convergence map that depends on the cosmology $\theta$ and the latent variables $z$.
    Given that the observational noise is uncorrelated across tomographic redshift bins and pixels, we can express the log-likelihood of the observed data $x_0$ as
    \begin{equation}\label{eq:logproba}
        \mathcal{L}(\theta, z) = constant - \sum_i^{N_{pix}} \sum_{j}^{N_{bins}} \frac{\left[\kappa_{ln}^{i,j}(\theta, z)-x_0^{i,j}\right]^2}{2\sigma_n^2}.
    \end{equation}
    By construction, $p(z | \theta)$ is independent of the cosmology $\theta$; hence, the log posterior we aim to sample is 
    \begin{equation}
        \log p(\theta, z | x = x_0) \propto \mathcal{L}(\theta, z) + \log p(z) + \log p(\theta),
        \label{eq:logprob}
    \end{equation}
    with $p(z)$ being a reduced-centered Gaussian and $p(\theta)$ as in \autoref{tab:fid&prior}.
    We used an HMC scheme to sample Eq. \ref{eq:logprob}. Specifically, we used the No-U-Turn sampler (NUTS, \citealp{hoffman2011nouturn}) from \href{https://github.com/pyro-ppl/numpyro}{\texttt{NumPyro}} \citep{phan2019composable, bingham2019pyro}, which efficiently proposes new relevant samples using the derivatives of the distribution we sampled from, namely $\nabla_{\theta, z} \log p(\theta, z | x = x_0)$. 
    
\begin{figure}[h]
        \includegraphics[width=9.cm]{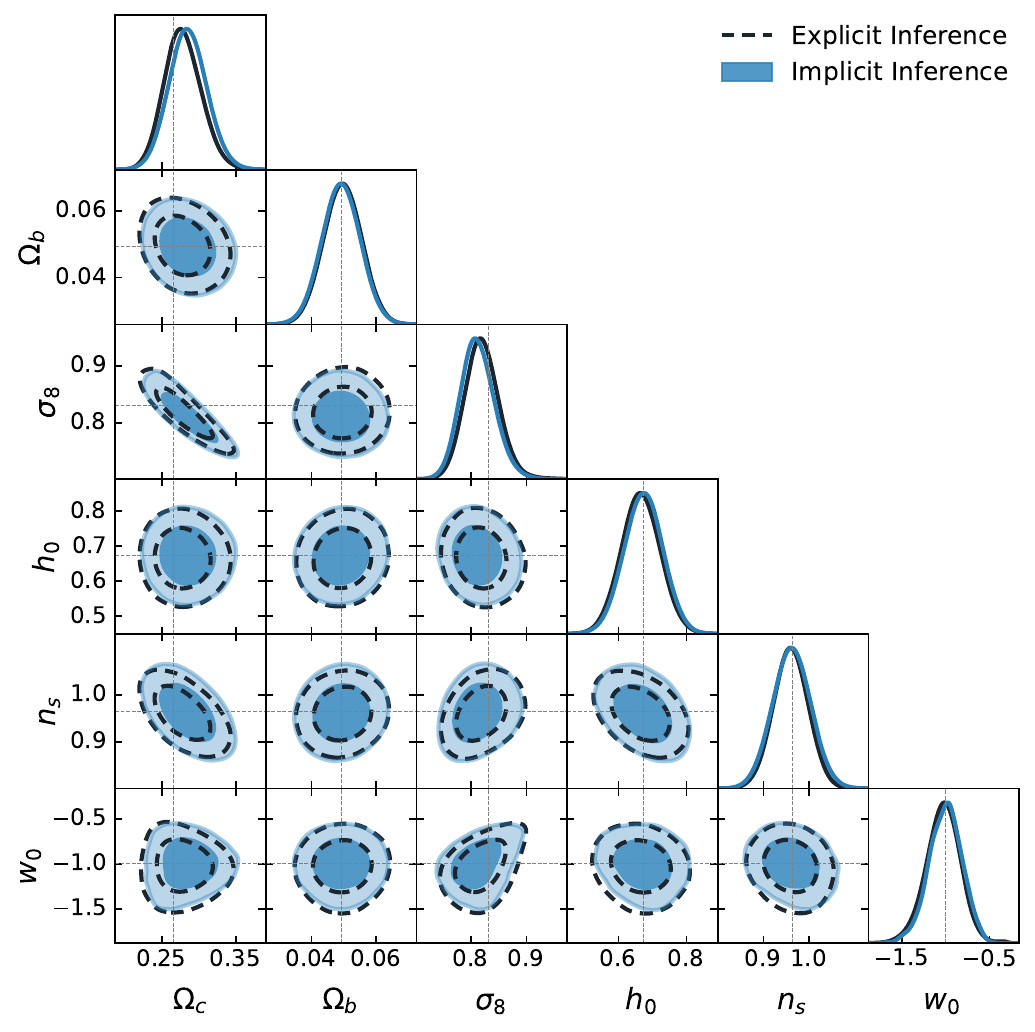}
        \caption{From log-normal simulated LSST Y10-like convergence maps, we constrain $w$CDM parameters using two approaches: 1) the explicit full-field inference (black), obtained by sampling $160\,000$ posterior samples through an HMC scheme; and 2) the implicit full-field inference contours (blue), obtained by compressing the convergence maps into sufficient statistics using variation mutual information maximization (VMIM) and performing inference using NLE with $1\,000$ simulations.
        We show three things: 1) implicit and explicit full-field inferences yield consistent constraints; 2) our implicit inference, when combined with an optimal compression procedure, allows full-field inference; 3) the C2ST metric indicates convergence when it is equal to $0.5$ (see Sect. \ref{sec:metric}). However, when comparing explicit and implicit inference (which should theoretically yield the same posterior), we never reach this value, but rather we obtain $0.6$. We justify that this value is acceptable and use it as a threshold for all benchmarked methods in this paper by showing that the marginals of the two approaches match, even though their C2ST is $0.6$.}
        \label{fig:conv_contour}
    \end{figure}

    Given the fixed observed convergence map $x_0$, Fig. \ref{fig:conv_contour} shows the posterior constraints on $\Omega_c, \Omega_b, \sigma_8, n_s, w_0, h_0$. As explained in Sect. \ref{sec:metric}, we consider this posterior of $160\,000$ samples converged as it yields the same constraint as our implicit full-field approach. Therefore, we consider these $160\,000$ samples as our ground truth.
    
    \subsection{Number of simulations for explicit full-field inference}
    We conducted a study to access the minimum number of simulations needed to achieve a good approximation of posterior distribution $p(\theta | x=x_0)$. In other words, we tried to access the minimum number of simulations needed to have converged MCMC chains and a good representation of the posterior distribution.
    
    Since there is no robust metric to estimate the convergence of MCMCs, and because we aim to compare all inference methods with the same metric, we used the C2ST metric to access the minimum number of simulations required to have converged chains. We proceed as follows: given the fully converged chains of $160\,000$ posterior samples from Fig. \ref{fig:conv_contour}, for each number of simulation $N$, we took the first $N$ samples and computed the C2ST metric comparing those samples to the $160\,000$ ones from the fully converged chains. The C2ST metric is based on the training of a classifier to distinguish between two populations under the cross-entropy loss and thus requires an equal number of samples of the two distributions. 
    We used a kernel density estimator (KDE) \citep{kde} to fit the samples, enabling us to generate the required number of samples to compare the two distributions. We note that KDEs, Gaussian filters, or smoothing are always used to visualize distribution samples using contour plots, thus motivating our approach. In addition, the distribution of interest is a six-dimensional unimodal and almost Gaussian distribution, making it easy to fit through a KDE. We used a Gaussian kernel and adjusted the bandwidth to align with the contour plots shown by GetDist, as highlighted in Fig. \ref{fig:kde_proof}.

    We note that samples and simulations are not the same thing. During each step, the proposal of the MCMC suggests a pair of parameters $\theta_1$ and $z_1$ and produces a corresponding simulation $x \sim p(x| \theta = \theta_1, z=z_1)$. The MCMC keeps only the parameters $\theta_1$ and $z_1$ if it yields a plausible $x$ according to the likelihood function evaluated on the observation $x_0$, and plausible $\theta_1$ and $z_1$ according to their priors. The sample is the pair $\theta$ and $z$ that is kept by the MCMC.
    Specifically, to obtain one posterior sample using the NUTS algorithm, we need $2 \times N$ simulations, with $N$ denoting the number of leapfrog steps. Indeed, the proposal of HMC methods is based on Hamiltonian equations that are discretized using the leapfrog integrator: 
    \begin{align}
        r^{\: t + \epsilon / 2} &= r^{\: t} - (\epsilon / 2) \nabla_{\alpha} \log p(\alpha^{\:t} | x_0),\\
        \alpha^{\: t + \epsilon } &= \alpha^{\:t} + \epsilon M^{-1}r^{\: t+\epsilon / 2},\\
        r^{\: t + \epsilon} &= r^{\: t + \epsilon / 2} -  (\epsilon / 2)\nabla_{\alpha} \log p(\alpha^{\: t+\epsilon} | x_0), 
    \end{align}
with $\epsilon$ being the step size and $M$ the mass matrix, $\alpha^{\: t}$ corresponding to the position of $(\theta, z)$ at time $t$, and $r^{\:t}$ denoting the values of the random momentum at time $t \in [0, N]$. After $N$ leapfrog steps, the total number of log probability evaluations is $N$. As each gradient requires the cost of two simulations (one to evaluate the primal values during the forward pass and one to evaluate gradients backward in the reverse mode of automatic differentiation), the total number of simulations is $2 \times N$. In our case, we find that the NUTS algorithm requires $2\times(2^6 -1) = 126$ simulations (always reaching the maximum depth of the tree that is set to $6$) to generate one sample.
    
\begin{figure}[h]
        \begin{center}
            \includegraphics[width=9.cm]{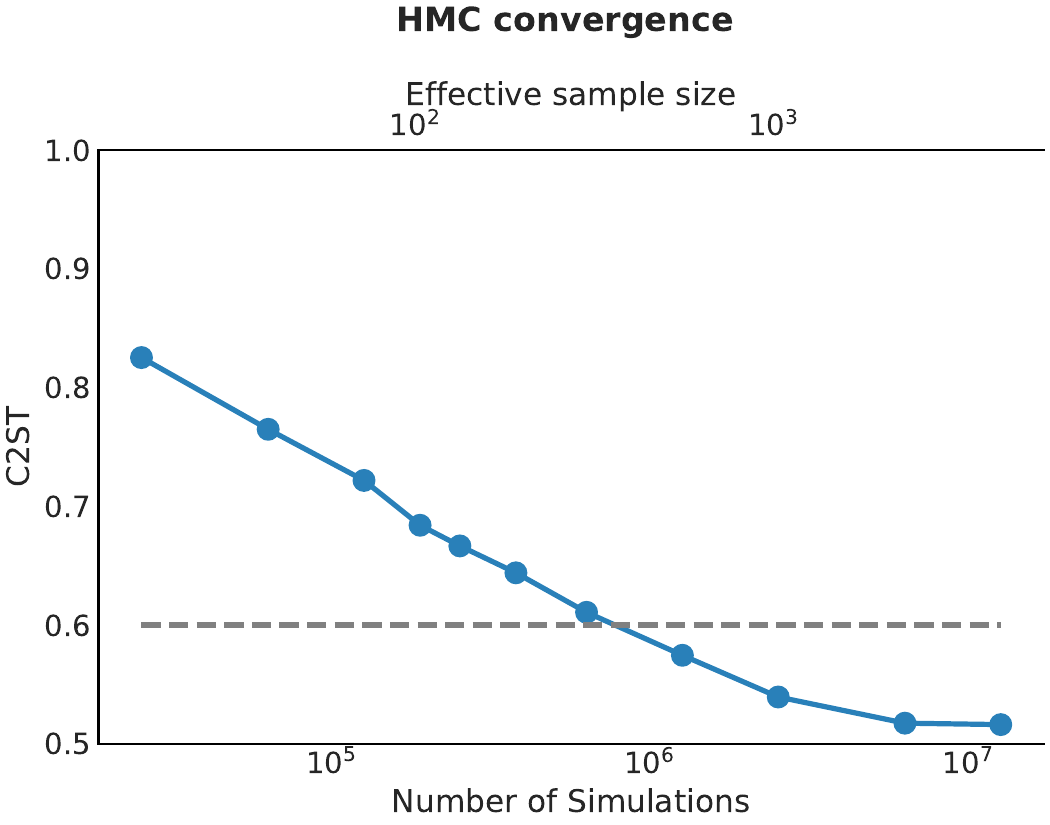}
            \caption{Explicit full-field inference: quality of cosmological posterior approximation as a function of the number of simulations used and effective sample size. The dashed line indicates the C2ST threshold of $0.6$, marking the point at which the posterior is considered equal to the true distribution (see Sect. \ref{sec:metric}).}
        \label{fig:hmc_cvg}
        \end{center}
    \end{figure}

    Figure \ref{fig:hmc_cvg} shows the convergence results of our explicit full-field inference as a function of the number of simulations and the effective sample size. According to the threshold of C2ST$= 0.6$ that we mention in Sect. \ref{sec:metric}, this study suggests that $630\,000$ simulations for our sampler corresponding to $400$ independent samples are enough to have converged MCMC chains. The number of independent samples is estimated using the effective-sample-size (ess) lower-bound estimate from \href{https://github.com/tensorflow/probability}{TensorFlow Probability} \citep{dillon2017tensorflow}. 
    In addition, Fig. \ref{fig:ei_contour_plots} shows the explicit posterior constraints obtained for different simulation budgets, and Fig. \ref{fig:ei_mean_std} shows the evolution of the mean and standard deviation of the posteriors as a number of simulations. We note that the C2ST metric is sensitive to higher order correlations, but if one only cares about marginals, the explicit inference posterior can be considered converged with only $63\,000$ simulations (corresponding to $24$ independent samples), as shown by the combination of contour plots Figs. \ref{fig:ei_contour_plots} and \ref{fig:ei_mean_std}.

    These results are not a strong statement about explicit inference in general as we do not investigate other sampling schemes and preconditioning schemes (this study is left for future work). However, the NUTS algorithm is one of the state-of-the-art samplers and has already been used in various full-field studies \citep[e.g.][]{zhou2023accurate, boruah2024bayesian}. However, it is important to note that there are other powerful HMC schemes, such as the Microcanonical Langevin Monte Carlo (MCLMC) method \citep{robnik2023microcanonical}, which might perform well with fewer simulations and has also been used in full-field studies \citep{bayer2023fieldlevel}. Regardless of the sampling scheme used, we suggest that readers refer to the effective sample size values to translate the results to their sampler.
    
\section{Implicit inference}

\label{sec:implicit_inference}
    Although explicit full-field inference offers a promising framework for performing rigorous Bayesian inference, it comes with the downside of requiring an explicit likelihood. Additionally, sampling from the joint likelihood even with HMC schemes can be very challenging and requires a large number of simulations. Instead, implicit inference has emerged as a solution to tackle the inference problem without relying on an explicit likelihood.
    These techniques rely on implicit likelihoods, which are more commonly known as simulators. A simulator is a stochastic process that takes as input the parameter space $\theta \sim p(\theta)$ and returns a random simulation $x$. It does not require the latent process of the simulator to be explicit. 
    
    Comparably to sampling the forward model, given an observation $x_0$, 
    one can simulate a large range of $\theta_i$ and accept the parameters that verify $|x_i - x_0| < \epsilon$ with $\epsilon$ a fixed threshold to build the posterior $p(\theta  | x=x_0)$. 
    This is the idea behind the approximate Bayesian computation (ABC) method 
    (e.g., \citealp{10.1214/aos/1176346785, 10.1093/genetics/162.4.2025, sisson2018overview}). 
    This method used to be the traditional way to carry out implicit inference, but its 
    poor scalability with dimension encouraged the community to develop new techniques. In particular, the introduction of machine learning leading to neural implicit inference methods has been shown to perform better.
    These neural-based methods cast the inference problem into an optimization task, 
    where the goal is to find the set of parameters $\varphi$ so that the neural parametric 
    model best describes the data. Then, the posterior is approximated using this 
    surrogate model evaluated on the given observation. 

    Implicit inference has already been successfully applied to cosmic shear analyses. For instance, \cite{Lin_2023} and \cite{vonwietersheimkramsta2024kidssbisimulationbasedinferenceanalysis} applied it to two-point statistics rather than using the standard explicit inference method assuming a Gaussian likelihood. Similarly, to bypass this traditional Gaussian likelihood assumption, \cite{Jeffrey2024} applied implicit inference to the power spectra, peak counts, and neural summary statistics.

    In this section, we introduce the NLE method and its augmented version with gradient $\partial$NLE, and we present the benchmark results. The neural ratio estimation (NRE), neural posterior estimation (NPE), and sequential methods are described in Appendix \ref{appendix:sbi_methods}; and the benchmark results of (S)NLE, (S)NPE, and (S)NRE can be found in Appendix \ref{appendix:sbi_results}.
    In this section, we focus our study on the NLE method as our comparison of the three main implicit inference methods (see Fig. \ref{fig:results_bm}) suggests that NLE and NPE are the ones that perform the best. We chose not to use the NPE method, as the augmented gradient version of NPE \citep{zeghal2022} requires specific neural architectures that proved to be more costly simulation-wise.

      \subsection{Learning the likelihood}
         The aim of the NLE method is to learn the marginal likelihood
         $p_{\varphi}(x | \theta)$ from a set of parameters and corresponding 
         simulations $(\theta, x)_{i=1..N}$. Thanks to the development 
         of new architectures in the neural density estimator field, this can be 
         achieved by using conditional normalizing flows (NFs) \citep{nfs}.
         Conditional NFs are parametric models $p_{\varphi}$ that take
         $(\theta, x)$ as input  and return a probability density $p_{\varphi}(x | \theta)$, 
         which can be evaluated and/or sampled.
         To find the optimal parameters $\hat{\varphi}$ that make
         $p_{\varphi}(x | \theta)$ best describe the data, one trains the NF
         so that the approximate distribution $p_{\varphi}(x | \theta)$ is
         the closest to the unknown distribution $p(x | \theta)$. To quantify this, we used the forward Kullback–Leibler divergence $D_{KL}(. \: || \: .)$. The $D_{KL}$ is positive and equal to zero if and only if the two distributions are the same, motivating the following optimization scheme:
         \begin{align}
            \hat{\varphi} &=  \arg \min_{\varphi} \mathbb{E}_{p(\theta)} \left [ D_{KL}(p(x | \theta) \: || \: p_{\varphi}(x | \theta)) \right ]  \\
            & = \arg \min_{\varphi} \mathbb{E}_{p(\theta)} \left [\mathbb{E}_{p(x | \theta)}\Big[ \log\left(\frac{p(x | \theta)}{p_{\varphi}(x | \theta)}\right) \Big] \right ]  \nonumber\\
            \begin{split}
            & = \arg \min_{\varphi} \underbrace{\mathbb{E}_{p(\theta)} \left [\mathbb{E}_{p(x | \theta)}\left[ \log\left(p(x | \theta)\right) \right] \right ]}_{\text{constant w.r.t }\varphi} \\
            &\qquad - \mathbb{E}_{p(\theta)} \left [\mathbb{E}_{p(x | \theta)}\left[ \log\left(p_{\varphi}(x | \theta)\right) \right] \right ]
            \end{split} \nonumber \\
            & =  \arg \min_{\varphi} - \mathbb{E}_{p(\theta)} \left [\mathbb{E}_{p(x | \theta)}\left[ \log\left(p_{\varphi}(x | \theta)\right) \right] \right ] \nonumber
        \end{align},
         leading to the loss
         \begin{equation}
            \mathcal{L} = \mathbb{E}_{p(\theta, x)} \left[ - \log p_{\varphi} (x | \theta)  \right],
            \label{eq:nll}
         \end{equation}
which does not require evaluation of the true target distribution $p(x | \theta)$ anymore. To compute this loss, only a set of simulations $(\theta, x) \sim p(\theta, x)$ obtained by first generating parameters from the prior $\theta_i \sim p(\theta)$ and then generating the corresponding simulation $x_i \sim p(x|\theta = \theta_i)$ through the simulator are needed. We note that the approximated likelihood, under the loss of Eq. \ref{eq:nll}, is learned for every combination $(\theta, x) \sim p(x, \theta)$ at once.

         Given observed data $x_0$, the approximated posterior 
         $\hat{p}(\theta | x=x_0) \propto p_{\hat{\varphi}}(x=x_0 | \theta) \: p(\theta)$ is then obtained by using an MCMC with the following log probability: $\log p_{\hat{\varphi}}(x=x_0 | \theta)+ \log p(\theta)$. This MCMC step makes NLE (and NRE) less amortized and slower than the NPE method, which directly learned the posterior distribution $p_\varphi(\theta | x)$ for every pair $(\theta, x) \sim p(x, \theta)$ and only needs to be evaluated on the desire observation $x_0$ to obtain the approximated posterior $p_\varphi(\theta | x = x_0)$. However, it is less challenging than using an MCMC scheme to sample the forward model in the explicit inference framework. Indeed, now one only has to sample the learned marginal likelihood $p_{\varphi}(x | \theta)$ (or learned likelihood ratio), not the joint likelihood of the forward model $p(x | \theta, z)$.

    \subsection{NLE augmented with gradients}
    \label{sec:grad_nle}
    
    Although there are methods to reduce the number of simulations, such as sequential approaches (see Appendix \ref{appendix:sbi}), they still treat the simulator as a black box. As underlined by \cite{Cranmer_2020}, the emergence of probabilistic programming languages makes it easier to open this black box (making the implicit likelihood explicit) and extract additional information such as the gradient of the simulation. In particular, \cite{mining_gold} noticed that they can compute the joint score $\nabla_{\theta} \log p(x,z | \theta)$ as the sum of the scores of all the latent transformations encountered in the differentiable simulator: 
        \begin{align}
           \nabla_{\theta} \log p(x,z | \theta) &=  \nabla_{\theta} \log p(x | \theta, z) +  \nabla_{\theta} \log p(z | \theta) \\
           &=  \nabla_{\theta} \log p(x | \theta, z) +  \sum_i^N \nabla_{\theta} \log p(z_i | z_1 ... z_{i -1}, \theta). 
        \end{align}
        The most important result is that through the use of the classical mean-squared error (MSE) loss (also known as score matching (SM) loss), 
        \begin{align}
         \label{eq:smloss}
             \mathcal{L}_{\rm SM} = \mathbb{E}_{p(x,z,\theta)} \left[ \parallel \nabla_{\theta} \log p(x,z | \theta)  - \nabla_\theta \log p_\varphi(x | \theta) \parallel_2^2 \right],
         \end{align}
        they showed how to link this joint score to the intractable marginal score $\nabla_{\theta} \log p(x \: | \theta)$. As explained in Appendix \ref{sec:demo_mse}, $\mathcal{L}_{\rm SM}$ is minimized by 
        $\mathbb{E}_{p(z |x,\theta)} \left[   \nabla_{\theta} \: \log \:  p(x, z |\theta) \right]$
        and can be derived as
        \begin{align}
            & \mathbb{E}_{p(z |x,\theta)} \left[   \nabla_{\theta} \: \log \:  p(x, z |\theta) \right] \nonumber\\
              & = \mathbb{E}_{p(z |x,\theta)} \left[   \nabla_{\theta} \: \log \: p(z|x , \theta) \right]  +  \nabla_{\theta} \: \log \:  p(x | \theta) \nonumber\\
              & = \mathbb{E}_{p(z| x,\theta)} \left[   \frac{ \nabla_{\theta}  \: p(z|x , \theta)} {p(z|x , \theta)} \right]  +  \nabla_{\theta} \: \log \:  p(x | \theta) \nonumber\\
              & = \int \nabla_{\theta}  \: p(z|x , \theta) \: dz  +  \nabla_{\theta} \: \log \:  p(x | \theta) \nonumber\\
              & =  \nabla_{\theta}  \: \int p(z|x , \theta) \: dz  +  \nabla_{\theta} \: \log \:  p(x | \theta) \nonumber\\
               & =   \nabla_{\theta} \: \log \:  p(x | \theta).
         \end{align}
        This loss learns how the probability of $x$ given $\theta$ changes according to $\theta$ and thus can be combined with the traditional negative log-likelihood loss (Eq. \ref{eq:marginal_likelihood}) to help the neural density estimator learn the marginal likelihood with fewer simulations. The NF now learns $p_{\varphi}(x | \theta)$ from $(\theta, x, \nabla_{\theta} \log p (x,z | \theta))_{i = 1..N}$ under the combined loss,
        \begin{equation}
             \mathcal{L} = \mathcal{L}_{\rm NLL} + \lambda  \: \mathcal{L}_{\rm SM},
         \end{equation}
         with $\lambda$ being a hyper-parameter that has to be fined-tuned according to the task at hand.
        \cite{mining_gold} called this method the SCore-Augmented Neural Density Approximates Likelihood (SCANDAL); we choose to rename it  $\partial$NLE for clarity in our paper. Equivalently, other quantities such as the joint likelihood ratio $r(x,z | \theta_0, \theta_1) = p(x,z | \theta_0) / p(x,z | \theta_1)$ and the joint posterior gradients $\nabla_{\theta} \log p(\theta | x, z)$ can be used to help learn the likelihood ratio \citep{mining_gold} and the posterior \citep{zeghal2022} respectively.

\subsection{Compression procedure}
\label{sec:compression}
In this section, we provide a brief summary of the compression procedure we performed to build sufficient statistics. A more detailed description and comparison of compression procedures applied in the context of weak-lensing, full-field implicit inference can be found in \cite{lanzieri2024optimalneuralsummarisationfullfield}.  

Based on the benchmark results of \cite{lanzieri2024optimalneuralsummarisationfullfield}, we chose to use the variational mutual information maximization (VMIM,  \citealp{jeffrey2021likelihood}) neural compression. This compression builds summary statistics $t = F_{\varphi}(x)$ by 
maximizing the mutual information $I(t, \theta)$ between the parameters of interest $\theta$ and the summary statistics $t$. More precisely, the mutual information is defined as
\begin{align}
    I(t, \theta) = \mathbb{E}_{p(t, \theta)} [\log{p(\theta | t )}]- H(\theta),
\end{align}
where $H$ denotes the entropy. Replacing the summary statistics $t$ by the neural network $F_\varphi$ and the intractable posterior $p(\theta| t)$ by a variational distribution $p_\psi(\theta| t)$ to be optimized jointly with the compressor, we obtain the following variational lower bound \citep{barber2003information}:
\begin{align}
    I(t, \theta) \geq \mathbb{E}_{p(x, \theta)} [\log{p_\psi(\theta \:  | \:F_\varphi(x) )}]- H(\theta).
\end{align}
Hence, by training the neural network $F_{\varphi}$ jointly with a variational distribution (typically a NF) $p_\psi$ under the loss 
\begin{equation}
    \mathcal{L}_{\text{VMIM}} = - \mathbb{E}_{p(x, \theta)} [\log{p_\psi(\theta \:  | \:F_\varphi(x) )}], 
\end{equation}
we were able, by construction and within the limit of the flexibility of $F_\varphi$ and $p_\psi$, to build summary statistics $t$ that contain the maximum amount of information regarding $\theta$ that is embedded in the data $x$. As equality is approached, the maximization of the mutual information yields sufficient statistics such that  $p(\theta | x) = p(\theta | t)$.

As proof that in our particular case, these summary statistics extract all the information embedded in our convergence maps, and thus are sufficient statistics, we show in Fig. \ref{fig:conv_contour} that the contours obtained using this compression and the NLE implicit inference technique allowed us to recover the explicit full-field constraints.

We used $100\,000$ simulations for the compression part and did not investigate the question of the minimum number of simulations required. 
Although we used a large number of simulations to train our compressor, we can produce near-optimal summary statistics without training a neural network, which eliminates the need for additional simulations. As an example,  \citet{10.1093/mnras/staa3165} showed that they can produce summary statistics using scattering transforms that result in constraints similar to those obtained by building summary statistics using a CNN trained under mean-absolute-error (MAE) loss. While it is not guaranteed that these scattering transform coefficients will provide sufficient statistics to perform full-field inference, we hope that advances in transfer learning will allow us to propose new compression schemes that need very few simulations. This is left for future work. Details regarding our compressor architecture can be found in 
Appendix \ref{appendix:compressor_arch}.

\subsection{Results}

      \begin{figure}[h!]
    \centering
    \includegraphics[width=9.cm]{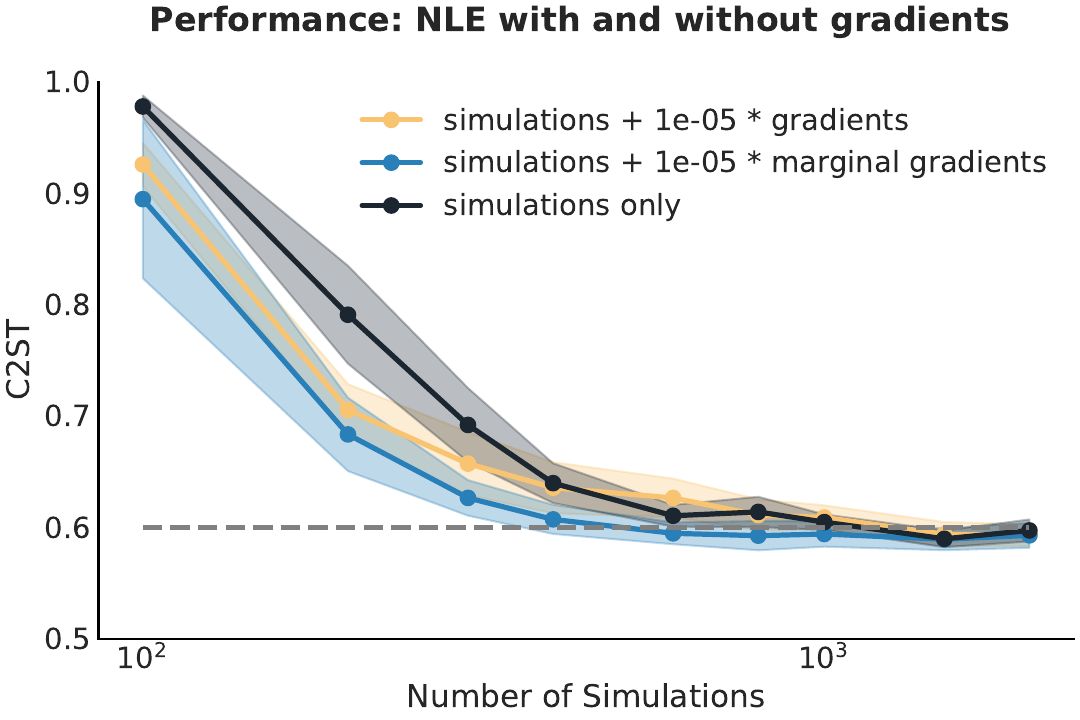}
    \caption{Implicit inference augmented with gradients: quality of the cosmological posterior approximation as a function of the number of simulations used. We compare three methods: 1) $\partial$NLE with the gradients of the simulator $\nabla_{\theta} \log p(x,z | \theta)$ (yellow); 2) $\partial$NLE with marginal gradients $\nabla_{\theta} \log p(x | \theta)$ (blue); and 3) the classical NLE method (black). The dashed line indicates the C2ST threshold of $0.6$, marking the point at which the posterior is considered equal to the true distribution (see Sect. \ref{sec:metric}). We show that the gradients provided by the simulator (yellow curve) do not help reduce the number of simulations as they are too noisy (see Fig. \ref{fig:stochasticiy_grads}). }
    \label{fig:cvg_grad}
\end{figure}
\begin{figure*}[!h]
    \centering
    \begin{minipage}{6.cm}
        \includegraphics[width=5.9cm]{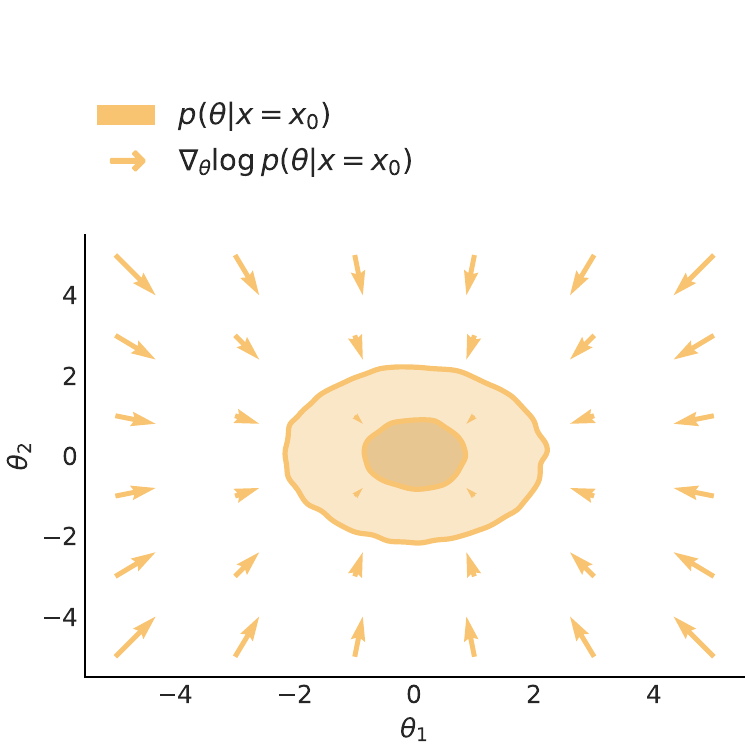}
    \end{minipage}
    \begin{minipage}{6.cm}
        \includegraphics[width=5.9cm]{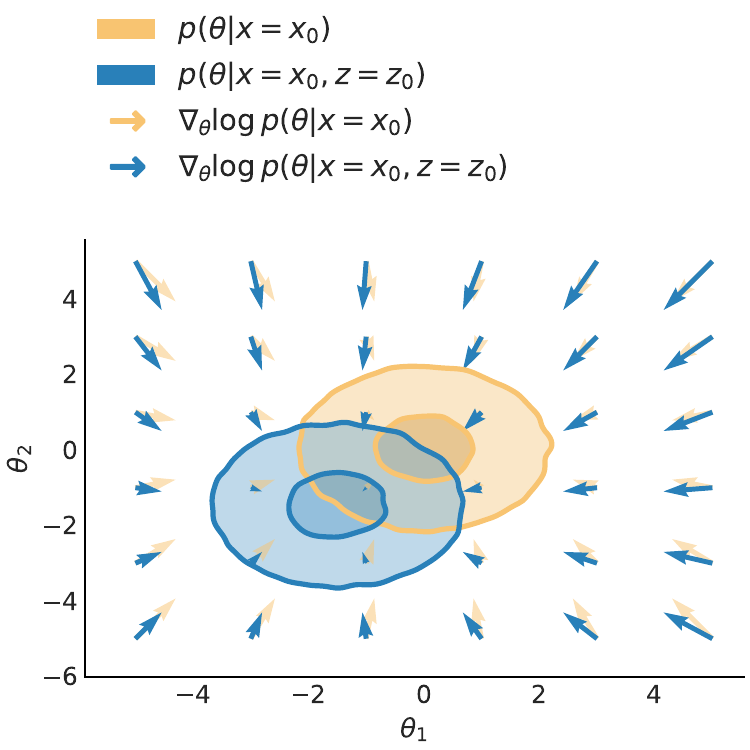}
    \end{minipage}
    \begin{minipage}{6.cm}
        \includegraphics[width=5.9cm]{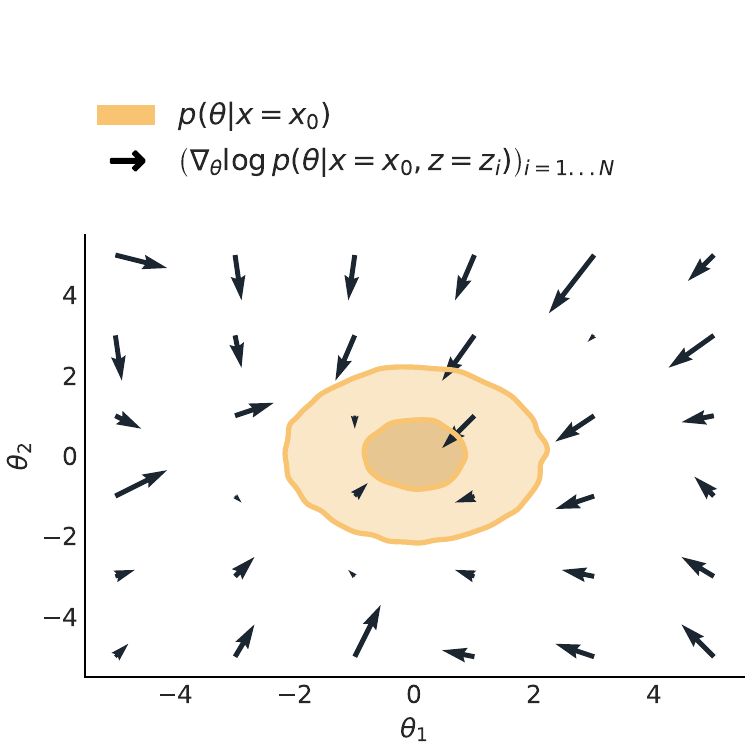}
    \end{minipage}
    \caption{Illustration of gradient stochasticity. The left panel shows a $2$D posterior distribution $p(\theta | x=x_0)$ evaluated at the observed data $x_0$ and its gradients $\nabla_{\theta} \log p(\theta | x=x_0)$. The middle panel shows the difference between the posterior $p(\theta | x=x_0)$ (yellow) and the joint posterior $p(\theta | x=x_0,z=z_0)$ (blue) with $z_0$ a latent variable that leads to $x_0$. The yellow arrows correspond to the gradients of the posterior, and the blue ones to the gradients of the joint posterior. The right panel displays the gradient field that we obtained in practice from the simulator. Each gradient aligns with its corresponding joint posterior, resulting in a "noisy" gradient field compared to that of the posterior (first panel).}
    \label{fig:stochasticiy_grads}
\end{figure*}

For this study, we used the NLE algorithm as in \cite{nle1}. And use the $\partial$NLE method introduced by \cite{mining_gold} to leverage gradient information. All approaches share the same NF architecture and sampling scheme (all details can be found in Appendix \ref{appendix:nn_arch_grad}).

We benchmark the previously presented implicit inference methods on our \sbilens's log-normal LSST Y10-like forward model. The goal of this inference problem is to constrain the following cosmological parameters: $\Omega_c, \Omega_b, \sigma_8, n_s,w_0, h_0$ given a fiducial convergence map $x_0$. Our fiducial map $x_0$ is the same for all the benchmarked methods in the paper.

This benchmark aims to find the inference method that can achieve a given posterior quality (C2ST = 0.6) with the minimum number of simulations; for this, the procedure is the following: 
\begin{enumerate}
    \item Starting from the entire dataset, we compress the tomographic convergence maps $x$ of $256 \times 256 \times 5$ pixels into  six-dimensional sufficient statistics. We use the VMIM  neural compression as described in Sect. \ref{sec:compression}.
    \item From this compressed dataset, we then pick a number of simulations and approximate the posterior distribution using  NLE and $\partial$NLE methods. 
    \item Then, we evaluate the approximated posterior against the fully converged explicit full-field posterior (our ground truth) using the C2ST metric. 
\end{enumerate}

The C2ST convergence results are displayed in Fig. \ref{fig:cvg_grad}. In the appendix, we provide additional convergence results. Fig. \ref{fig:nle_contours_evol} shows the posterior contours' evolution obtained through NLE, and Figs. \ref{fig:nle_mean_std}, \ref{fig:partialnle_mean_std}, \ref{fig:marginalpartialnle_mean_std} depict the evolution of the mean and standard deviation of the approximated posterior as a number of simulations.

We find that unlike previous results \citep{mining_gold, zeghal2022}, the gradients $\nabla_{\theta} \log p(x,z | \theta)$ do not provide additional information enabling a reduction in the number of simulations. Indeed,  Fig. \ref{fig:cvg_grad} shows similar convergence curves for the $\partial$NLE method (yellow) and the NLE method (black). 
This issue arises as we attempt to constrain the gradients of the learned marginal distribution $p_{\varphi}(x | \theta)$ by using the joint gradients  $\nabla_{\theta} \log p(x,z | \theta)$ from the simulator. Indeed, the benefit of these joint gradients depends on their "level of noise". In other words, their benefit depends on how much they vary compared to the marginal gradients. To visually exhibit this gradient stochasticity, we considered the gradients of a two-dimensional posterior $\nabla_{\theta} \log p(\theta | x)$ and the joint gradients $\nabla_{\theta} \log p(\theta | x, z)$ provided by the simulator. By definition, the gradients should align with the distribution, as seen in the left panel of  Fig. \ref{fig:stochasticiy_grads}. As demonstrated in the middle panel of Fig. \ref{fig:stochasticiy_grads}, the gradients we obtain from the simulator are directed toward $p(\theta | x, z)$, which differs from $p(\theta | x) = \int p(\theta | x, z)p(z|x)dz$. The stochasticity of the gradients relies on the standard deviation of $p(z|x)$ and how much $p(\theta | x, z)$ "moves" according to $z$.
As a result, instead of the gradients field displayed in the left panel of  Fig. \ref{fig:stochasticiy_grads}, we end up with the gradients field depicted in the right panel.

To confirm this claim, we learn from the simulator's gradients, $\nabla_{\theta} \log p(x,z | \theta)$, and the marginal ones, $\nabla_{\theta} \log p(x | \theta)$. For this, we used a neural network (the architecture can be found in Appendix \ref{appendix:marginal_grad_arch}) that we trained under the following MSE loss function: 
\begin{align}
     \mathcal{L}_{\rm MarginalSM} = \mathbb{E}_{p(x,z,\theta)} \left[ \parallel \nabla_{\theta} \log p(x,z | \theta)  - g_\varphi(x,\theta) \parallel_2^2 \right].
\end{align}
This loss is almost the same as Eq. \ref{eq:smloss}, except that instead of using an NF to approximate $\nabla_{\theta} \log p(x | \theta)$ and then take its gradients, we trained a neural network to approximate the gradient values given $\theta$ and $x$. This loss is minimized by $g_\varphi(x,\theta) = \nabla_{\theta} \log p(x | \theta)$ (as explained in Sect. \ref{sec:grad_nle}), allowing us to learn the intractable marginal gradients from simulations. 

We then used these marginal gradients in the $\partial$NLE method (blue curve) and show that those gradients help reduce the number of simulations. What we demonstrate here is that the stochasticity of our LSST Y10-like simulator dominates the gradient information, and thus the $\partial$NLE method does not help perform inference with fewer simulations. 

We could have used a method to denoise the gradients. Specifically, \citet{Millea_2022} introduced marginal unbiased score expansion (MUSE), a way of computing marginal gradients from simulations, and proposed a frequentist and Bayesian approach for parameter inference that leverages this quantity. In our case, the $\partial$NLE with marginal gradients converges with $\sim 400$ simulations, while $\partial$NLE with gradients from the simulator converges with twice as many simulations. Hence, to be beneficial, computing the marginal gradient should take fewer than two simulations, which is not feasible with MUSE as it requires at least ten simulations to have an "acceptable" estimation of the marginal gradient \citep{Millea_2022}.

\section{Conclusion and discussion}
\label{sec:conclusion}

Full-field inference is the optimal form of inference as it aims to perform inference without any loss of information. This kind of inference is based on a simulation model known as a simulator, forward model, or Bayesian hierarchical model in cases where the model is hierarchical.
There are two ways of conducting full-field inference from this forward model: through explicit or implicit inference. The first way can be applied when the forward model is explicit. This means that the field-based joint likelihood $p(x = x_0|\theta,z)$ can be evaluated and thus sampled through sampling schemes such as MCMC. The second one can be used when only simulations are available; in this case, it is said that the likelihood is implicit. While it is possible to perform implicit inference directly at the pixel level by feeding the maps to the neural density estimator \citep{dai2024multiscale}, it is usually more robust and safe to break it down into two steps: first performing a lossless compression, and second performing the implicit inference on these sufficient statistics. Specifically, in this work, sufficient statistics are built using an optimal neural-based compression based on the maximization of the mutual information $I(\theta, t)$ between the cosmological parameters $\theta$ and the summary statistics $t$. However, other compression schemes, requiring fewer or zero simulations, could be used while still offering very good quality summary statistics \citep{10.1093/mnras/staa3165}. Additionally, the advent of transfer learning could offer a way of performing compression with fewer simulations; this is left for future work.

This work aimed to find which full-field inference methods require the minimum number of simulations and if differentiability is useful for implicit full-field inference. To answer these questions, we introduced a benchmark that compares various methods to perform weak lensing full-field inference. For our benchmark, we used \sbilens's differentiable forward model, which generates log-normal convergence maps imitating a simplified version of the LSST Y10 quality. We evaluated the performance of several inference strategies by evaluating the constraints on $(\Omega_c, \Omega_b, \sigma_8, h_0, n_s, w_0)$, specifically using the C2ST metric.

We found the following results: 
\begin{enumerate}
    \item Explicit and implicit full-field inference yield the same constraints. However, according to the C2ST metric and the threshold of C2ST $ = 0.6$, the explicit full-field inference requires $630\,000$ simulations (corresponding to $400$ independent samples). In contrast, the implicit inference approach requires $101\,000$ simulations split into $100\,000$ simulations for compression and $1\,000$ for inference. We note that we arbitrarily used $100\,000$ simulations for the compression part and did not explore the question of performing compression with a minimum number of simulations. Hence, $101\,000$ simulations is an upper bound of the number of simulations actually required to perform implicit full-field inference in this particular problem. 
    \item The C2ST is sensitive to higher order correlations that one cannot see by looking at the marginals or first moments, making it a good metric for comparing distributions. However, as we are mostly interested in those marginals, it is worth noting that by looking at the combination of contour plots from Fig. \ref{fig:ei_contour_plots} and first-moments convergence plots from Fig. \ref{fig:ei_mean_std}, the explicit inference can be considered "converged" with $63\,000$ simulations (corresponding to $24$ independent samples), as emphasized by Fig. \ref{fig:ei_contour_plots}, which correspond to C2ST=$0.76$ and the implicit inference performed through NLE with $101\,000$ ($1\,000$ for inference and $100\,000$ to build sufficient statistics) as shown in Fig. \ref{fig:conv_contour} which corresponds to C2ST=$0.6$.
    \item For implicit inference, we exploited the simulator's gradient using the SCANDAL method proposed by \cite{mining_gold}. Our study indicates that the gradients contain a significant noise level due to the latent variable's behavior, which makes it difficult to achieve convergence with fewer simulations. We note that the effectiveness of such gradient-based methods depends on the specific problem at hand. These methods can still be useful in scenarios where the noise level is not significant. This has been demonstrated in studies such as \cite{mining_gold} and  \cite{zeghal2022}. It is also important to keep in mind that there may be other ways to leverage the differentiability of simulators and encourage further research in this area. Finally, we note that methods to denoise the gradients exist \citep{Millea_2022}, but, in our specific case, the gain compared to the number of simulations that this method requires is not significant.

\end{enumerate}

It is worth noting that for each explicit inference simulation budget, the C2ST is calculated against fully converged explicit inference samples, resulting in a value that can reach almost $0.5$. For implicit inference, the C2ST is also computed against the fully converged explicit inference samples. Both methods should produce the same constraints, but due to slight differences in the posterior approximation, the C2ST cannot go below $0.6$. Hence, we consider a value of $0.6$ as indicating convergence (see Fig. \ref{fig:conv_contour}). 

It is important to mention that in most real-world physical inference problems, such a metric cannot be used as it requires a comparison of the approximated posterior to the true one. Instead, for implicit inference, coverage tests \citep{tarp} should be used to assess the quality of the posterior. For explicit full-field inference, although diagnostics exist, it is very difficult to verify if the MCMC has explored the entire space. If possible, the safest option would be to compare the two full-field approaches, as they should yield the same posterior. Implicit inference is likely the easiest to use in such a scenario because it does not require the modeling of the very complicated latent process of the forward model and can be performed even in multimodal regimes; whereas, explicit inference has to sample the latent process of the forward model, and the more dimensions there are, the more time it needs to explore the entire parameter space. In addition, it can fail in the case of multimodal distribution as it can stay stuck in local maxima and never converge. However, for implicit inference, too few simulations can result in an overconfident posterior approximation, as shown in Fig. \ref{fig:nle_contours_evol}. Therefore, within the limit of a reasonable number of simulations, the implicit inference method should be the easiest to use.

Finally, we discuss several limitations of our approach. Although the current setup is not fully realistic for LSST, this minimalist approach was deliberately chosen to benchmark explicit and implicit inference within a manageable time frame. It is important to note that we used log-normal simulations instead of more computationally intensive N-body PM simulations. While this model accounts for additional non-Gaussianity (as illustrated in Fig. \ref{figappendix:lognormal}), it does not match the complexity of N-body simulations.
Modeling systematics at the field-level likelihood is inherently complex and poses challenges for implementing explicit inference. Similarly, incorporating other forms of noise and masking at the field level may be difficult. To ensure a fair comparison, we opted not to include these complexities, which enabled us to use the same simulator for both inference methods to maintain consistency across the comparison.
While it is true that an incorrect forward model would impact the results when applying this framework to real observations, we do not expect our conclusions to change regarding the comparison of implicit and explicit inference. However, it will be interesting to validate this assumption with a more realistic gravity model in future work. The results presented here should be viewed as illustrative benchmarks rather than definitive predictions for LSST data, though we remain optimistic that our findings will be relevant for realistic weak lensing inference.

The explicit inference results are not a strong statement, as we did not explore other sampling and preconditioning schemes (which is left for future work). Our sampler choice for the benchmark was motivated by the fact that the NUTS algorithm is a state-of-the-art sampler and has been extensively used in full-field studies \citep{zhou2023accurate, boruah2024bayesian}. However, there are other sampling schemes, such as powerful microcanonical Langevin Monte Carlo (MCLMC) methods \citep{robnik2023microcanonical} which might require fewer simulations and have been applied in full-field studies \citep{bayer2023fieldlevel}. Meanwhile, we recommend that the reader refer to the effective sample size values to translate the results to their sampler.

We used the NLE implicit method for our study as, regarding our benchmark results of Fig. \ref{fig:results_bm}, it seems to be the one that performs the best. NPE provides comparable results, but necessitates using the $\partial$NPE method of \citet{zeghal2022} to leverage gradient information. Since the NPE method aims to learn the posterior directly, this method requires the NF to be differentiable. However, the smooth NF architecture \citep{smoothnf} that \citet{zeghal2022} used was too simulation-expensive for our needs. We also experimented with continuous normalizing flows trained under negative log-likelihood loss, but found that it took a very long time to train. 
\\

\begin{acknowledgements}
This paper has undergone internal review in the LSST Dark Energy Science Collaboration. The authors would like to express their sincere gratitude to the internal reviewers, Alan Heavens and Adrian Bayer, for their valuable feedback, insightful comments, and suggestions, which helped to significantly improve the quality of this work. They also extend their thanks to Benjamin Remy for his contributions through countless discussions and helpful comments on the paper. Additionally, they appreciate the constructive feedback provided by Martin Kilbinger and Sacha Guerrini.

JZ led the project, contributed to brainstorming, developed the code, and wrote the paper.
DL contributed to brainstorming and code development, particularly in developing the forward model, and reviewed the paper.
FL initiated the project and contributed through mentoring, brainstorming, code development, and paper reviews.
AB contributed mentoring, brainstorming, code development, and paper reviews.
GL and EA provided mentoring, participated in brainstorming, and contributed to reviewing the paper. AB contributed to the review of the paper and participated in brainstorming the metric used for explicit inference.

The DESC acknowledges ongoing support from the Institut National de 
Physique Nucl\'eaire et de Physique des Particules in France; the 
Science \& Technology Facilities Council in the United Kingdom; and the
Department of Energy, the National Science Foundation, and the LSST 
Corporation in the United States.  DESC uses resources of the IN2P3 
Computing Center (CC-IN2P3--Lyon/Villeurbanne - France) funded by the 
Centre National de la Recherche Scientifique; the National Energy 
Research Scientific Computing Center, a DOE Office of Science User 
Facility supported by the Office of Science of the U.S.\ Department of
Energy under Contract No.\ DE-AC02-05CH11231; STFC DiRAC HPC Facilities, 
funded by UK BEIS National E-infrastructure capital grants; and the UK 
particle physics grid, supported by the GridPP Collaboration.  This 
work was performed in part under DOE Contract DE-AC02-76SF00515.

This work was supported by the Data Intelligence Institute of Paris (diiP), and IdEx Université de Paris (ANR-18-IDEX-0001). 

This work was granted access to the HPC/AI resources of IDRIS under the allocations 2023-AD010414029 and AD011014029R1 made by GENCI.

This work used the following packages: \texttt{Numpy} \citep{numpy}, \texttt{NumPyro} \citep{phan2019composable}, \texttt{JAX} \citep{jax2018github},  \texttt{Haiku} \citep{haiku2020github}, \texttt{Optax} \citep{deepmind2020jax}, \texttt{JAX-COSMO} \citep{jaxcosmo}, \texttt{GetDist} \citep{lewis2019getdist}, \texttt{Matplotlib} \citep{plt}, \texttt{CosMomentum} \citep{cosmomentum}, \texttt{scikit-learn} \citep{scikit-learn}, \texttt{TensorFlow} \citep{tensorflow2015-whitepaper}, \texttt{TensorFlow Probability} \citep{dillon2017tensorflow}, \texttt{sbi} \citep{tejero-cantero2020sbi} and \texttt{sbibm} \citep{lueckmann2021benchmarking}.

\end{acknowledgements}

\bibliographystyle{aa} 
\bibliography{biblio} 

\begin{appendix}
\section{Log-normal simulations}\label{appendix:lognormal}

The following plot demonstrates that log-normal simulations can mimic the non-Gaussian behavior of late-time fields. Indeed, the constraints obtained from the full-field approach (sampling the forward model) are much tighter compared to the standard power spectrum analysis.

\begin{figure}[h]
    \centering
    \includegraphics[width=9cm]{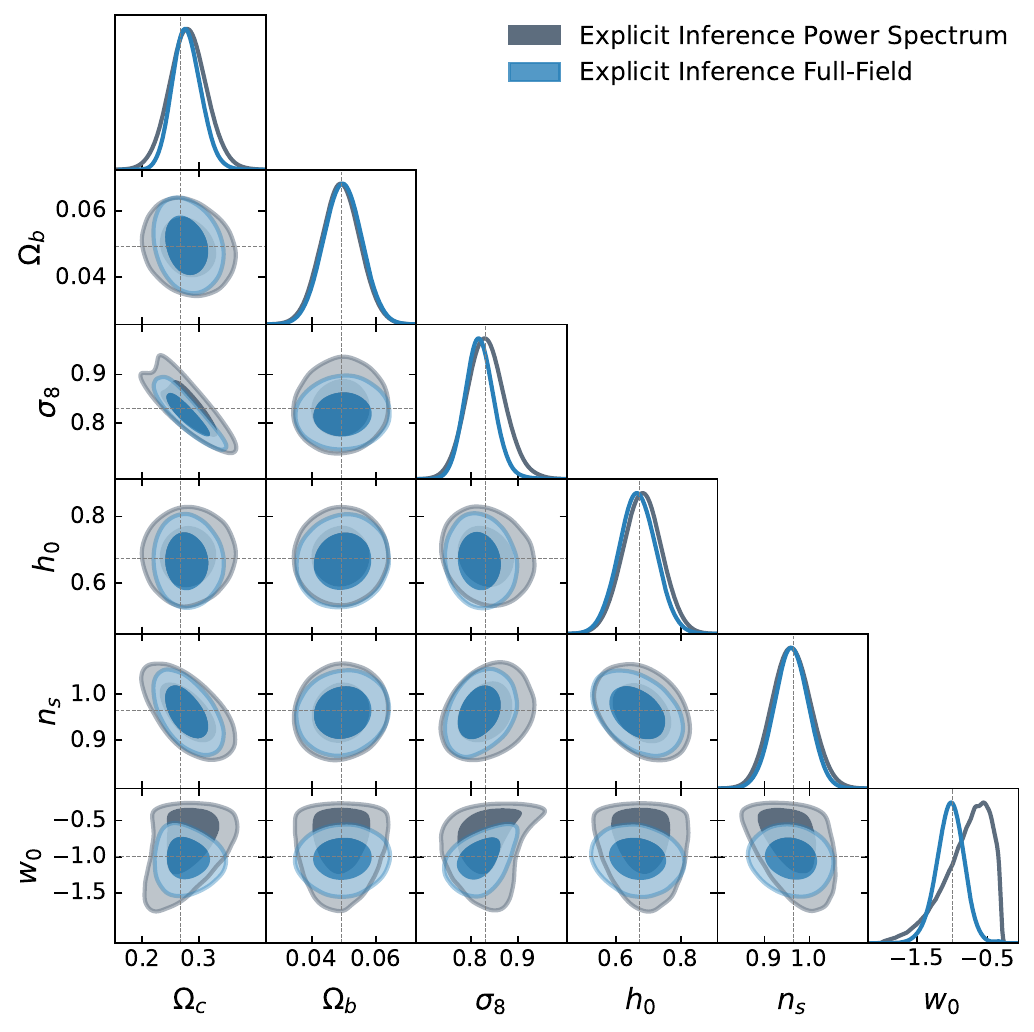}
    \caption{
    From log-normal simulated convergence maps at LSST Y10 quality we
    constrain the $w$CDM parameters using different inference techniques: power spectrum analysis (gray) and full-field analysis performed by sampling the forward model using an HMC sampler (blue). 
    }
    \label{figappendix:lognormal}
\end{figure}
\section{Implicit inference benchmark}
\label{appendix:sbi}
\subsection{Methods}
\label{appendix:sbi_methods}
\subsubsection{Learning the likelihood ratio}
Neural ratio estimation (NRE) is based on the well-known likelihood ratio test. The idea is to test whether $x$ has been generated by $\theta_0$ or $\theta_1$ through the following quantity: 
         \begin{equation}
            r(x | \theta_0, \theta_1) = \frac{p(x \: | \theta_0)}{p(x \: | \theta_1)}.
         \end{equation}
         Using the likelihood ratio trick this test can be cast as a binary  classification problem where we train a classifier $d_{\varphi}$ to learn the probability that $x$ has been generated by $\theta_0$: 
         \begin{align}
                d_{\varphi}(x) = p(y = 1 | x) = \frac{p(x \: | \theta_0) }{p(x |\theta_0) + p(x |\theta_1)}, 
         \end{align}
         \begin{equation}
            r(x | \theta_0, \theta_1) = \frac{d_{\varphi}(x)}{1 - d_{\varphi}(x)},
         \end{equation}
         with the two labels $y = 0$ and $y = 1$ corresponding respectively to $x \sim p(x | \theta_1)$ and $x \sim p(x | \theta_0)$.

         Finally, this is generalized to all possible parameters $\theta$ by defining the label $y = 0$ as $(x, \theta) \sim p(x) p(\theta)$ and the label $y = 1$ corresponding to $(x, \theta) \sim p(x, \theta)$. This means that now the classifier learns
         \begin{align}
                d_{\varphi}(x, \theta) = \frac{p(x, \theta) }{p(x) p(\theta) + p(x, \theta)} = \frac{p( \theta | x)}{p( \theta | x) + p(\theta)},
         \end{align}
        leading to the following likelihood ratio
         \begin{equation}
            r(x \: , \theta) = \frac{d_{\varphi}(x, \theta)}{1 - d_{\varphi}(x, \theta)} = \frac{p(\theta | x)}{p(\theta)}.
         \end{equation}

         \cite{lr4} generalized this binary classification into a $K$ multi-class classification and showed performance improvement when $K > 2$.
    
         Similarly to NLE, given observed data $x_0$, the approximated posterior is then obtained by sampling the distribution.
       
        \subsubsection{Learning the posterior}
         Neural posterior estimation (NPE) aims to directly 
         learn the posterior distribution. Similarly to NLE, 
         NPE is based on neural density estimators such as NFs, whose goal 
         is to learn $p_{\varphi}(\theta  | x)$ from a set of parameters 
         and corresponding simulations $(\theta, x)_{i=1..N}$. This can be done 
         by using a conditional NF and minimizing $D_{KL}$:
         \begin{align}
         \hat{\varphi} =  \arg \min_{\varphi}   \mathbb{E}_{p(x)} \left [ D_{KL}(p(\theta |x) \: || \: p_{\varphi}(\theta |x)) \right ],  \\
         \implies
            \mathcal{L} = \mathbb{E}_{p(\theta, x)} \left[ - \log p_{\varphi} (\theta |x)  \right].
            \label{eq:loss_npe}
         \end{align}
         Note that, unlike NLE and NRE, for NPE no MCMC is needed to get samples from the posterior. This approach is very convenient if one has to evaluate the posterior distribution for different observations as it only requires a new evaluation of the learned model $p_{\varphi}(\theta  | x)$. 

    \subsection{Sequentially refined posterior}

    In most cases the prior $p(\theta)$ is significantly broader compared to the posterior $p(\theta | x = x_0)$, making it unnecessary to sample the entire parameter space.
    Instead, we would like to sample from a proposal $\Tilde{p}(\theta)$ which 
    denotes the most suitable regions. The question arises: How to choose this proposal $\Tilde{p}(\theta)$ if we know neither the posterior location nor its size? 
    
    Starting from the prior, sequential methods offer a way to iteratively select this proposal by using the previous posterior approximation as the new relevant area and consequently refining the posterior at each iteration.

    Each of the methods described above (NPE, NLE, and NRE) can be sequentially adjustable, however, there are some specificities to bear in mind:
    both SNLE \citep{nle1} and SNRE \citep{lr4, hermans2020likelihoodfree} necessitate a sampling method or variational inference \citep{SNPLA, vi} at the end of each iteration to obtain the new parameters $\theta$. 
    SNPE \citep{npe1, npe2, npe3, deistler2022truncated} usually requires a costly correction of the approximated posterior since now minimizing the loss from Eq. \ref{eq:loss_npe} under the proposal $\Tilde{p}(\theta)$ leads to 
    \begin{equation}
          \tilde{p}(\theta |x) = p(\theta |x)  \frac{\tilde{p}(\theta) \: p(x)}{p(\theta) \: \tilde{p}(x)}. 
      \end{equation}
    
\subsection{Results}
      \label{appendix:sbi_results}

    \begin{figure*}[h]
        \begin{center}
        \includegraphics[width=\textwidth]{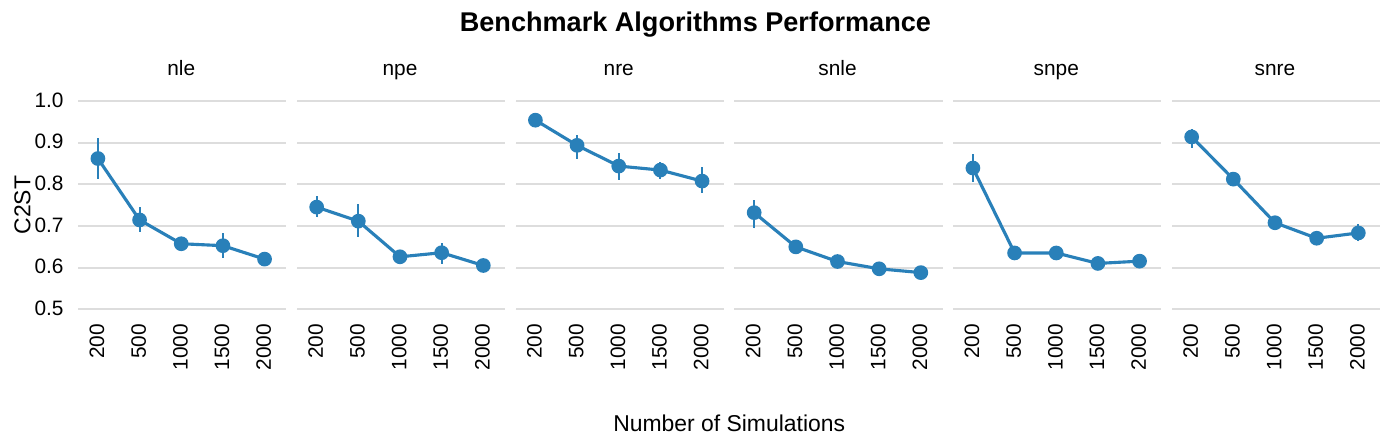}
        \caption{Implicit inference: quality of the cosmological posterior approximation as a function of the number of simulations used. We compare six methods using the default implementation of the \texttt{sbi} package: NLE, NPE, NRE, and their sequential counterparts SNLE, SNPE, and SNRE.}
        \label{fig:results_bm}
         \end{center}
    \end{figure*}

To benchmark (S)NLE, (S)NPE, and (S)NRE methods, we use the same benchmark procedure as the one presented in Sect. \ref{sec:implicit_inference}.
We use the \href{https://sbi-dev.github.io/sbi/}{\url{sbi}} package for (S)NPE, (S)NLE, and (S)NRE methods. We choose to rely on \texttt{sbi}'s developers' expertise and use the default setting of \texttt{sbi} but optimizing the architectures would be interesting future work. For now, more detail about the implementation of these algorithms can be found in Appendix \ref{appendix:standardsbi}.

Our numerical results in Fig. \ref{fig:results_bm}, suggest that NPE and NLE methods perform the best. The results also show that the sequential methods outperform their nonsequential analog. In particular, we find that SNPE and SNLE are the methods to favor as they allowed us to achieve a posterior quality of $0.6$ with only $1\,000$ simulations.

\section{MSE minimization}
\label{sec:demo_mse}
In this section we demonstrate that the following loss function
\begin{align*}
    \mathcal{L} &= \mathbb{E}_{p(x,z,\theta)} \left[ \parallel g(x, \theta, z)  - g_\varphi(x,\theta) \parallel_2^2 \right],
    \label{eq:fun}
\end{align*}
is minimized $\forall (x,\theta) \sim p(x,\theta)$ by 
\begin{align*}
 g_\varphi(x,\theta) = \mathbb{E}_{p(z | x,\theta)} \left[   g(x, \theta, z) \right].
\end{align*}
This proof is inspired by \citet{Remy2023}.

The optimal parameters of neural networks are typically chosen to cancel the following gradient:
\begin{align*}
    \frac{\partial \mathcal{L}}{\partial \varphi} &= \frac{\partial }{\partial \varphi} \: \mathbb{E}_{p(x,z,\theta)} \left[ \parallel g(x, \theta, z)  - g_\varphi(x,\theta) \parallel_2^2 \right]\\
    &= \frac{\partial }{\partial g_\varphi} \: \mathbb{E}_{p(x,z,\theta)} \left[ \parallel g(x, \theta, z)  - g_\varphi(x,\theta) \parallel_2^2 \right] \times \frac{\partial g_\varphi}{\partial \varphi}.
\end{align*}

\noindent Since $g_\varphi$ is by construction very unlikely to have null derivatives with respect to its parameters it means that 
\begin{align*}
    \frac{\partial }{\partial g_\varphi} \: \mathbb{E}_{p(x,z,\theta)} \left[ \parallel g(x, \theta, z)  - g_\varphi(x,\theta) \parallel_2^2 \right] =0.
\end{align*}

Thanks to Leibniz integral rule
we can switch the gradient and integrals such that 
\begin{align*}
\mathbb{E}_{p(x,z,\theta)} \left[ \frac{\partial }{\partial g_\varphi} \parallel g(x, \theta, z)  - g_\varphi(x,\theta) \parallel_2^2 \right] &= 0\\
\mathbb{E}_{p(x,z,\theta)} \left[  -2 \: g(x, \theta, z) + 2 \: g_{\varphi}(x,\theta)\right] &= 0\\
\mathbb{E}_{p( x,\theta)} \left [ -2  \mathbb{E}_{p(z| x,\theta)} \left[  \: g(x, \theta, z)\right] + 2 \: g_{\varphi}(x,\theta)\right] &= 0. 
\end{align*}

As $ \mathbb{E}_{p(x,z,\theta)} \left[\parallel g(x, \theta, z)  - g_\varphi(x,\theta) \parallel_2^2\right]$ is convex with respect to $g_\varphi$, it has a unique minimum that is reached when 
\begin{align*}
 g_\varphi(x,\theta) = \mathbb{E}_{p(z | x,\theta)} \left[   g(x, \theta, z) \right].
\end{align*}

\section{Experiments additional information}\label{appendix:nn_arch}

Codes for the compressor, the forward model, and the explicit full-field analysis are available at \href{https://github.com/DifferentiableUniverseInitiative/sbi_lens}{\sbilens}. All codes relative to the benchmark of implicit inference techniques are available at \href{https://github.com/LSSTDESC/sbi_bm_lens/tree/main}{\url{sbi\_bm\_lens}}.

\subsection{Compressor acrchitecture}\label{appendix:compressor_arch}
To compress the convergence maps of $5 \times 256 \times 256$ pixels into a $6$ dimensional summary statistics we used a residual neural network (ResNet) \citep{resnet} architecture. Specifically the ResNet-$18$. The ResNet-$18$ was trained under the VMIM loss function as described in Sect. \ref{sec:compression}.

\subsection{Neural network architecture to learn marginal gradients}\label{appendix:marginal_grad_arch}

To learn the marginal gradients $\nabla_{\theta}p(x \: | \theta)$ from the joint stochastic one $\nabla_{\theta}p(x, z\: | \theta)$ provided by the simulator, we used a neural network with $2$ layers of $256$ hidden units and Leaky ReLU activation functions. To test that we learned the correct marginal gradients we compared them against the gradients of a conditional NF trained with $10^5$ simulations under the NLE loss.

\subsection{NLE and $\partial NLE$ architectures}\label{appendix:nn_arch_grad}
For this study, the NF architecture remains fixed for the two methods, only the input changes: 1) we used only simulations; 2) we used simulations and the gradients of the simulator; 3) we used the simulations and the learned marginal gradients. Our conditional NF is a RealNVP \citep{dinh2017density}
of $4$ coupling layers. Scale and shift parameters are learned using a neural network of $2$ layers of $128$ hidden units each. We used SiLU activation functions. To get the posterior from the learned likelihood, we used NUTS sampler. 
The epistemic uncertainty is approximated by training $7$ NFs.

\subsection{Standard implicit inference architectures}\label{appendix:standardsbi}
To compare all the implicit inference techniques, we used the \href{https://sbi-dev.github.io/sbi/}{\url{sbi}} package for (S)NPE, (S)NLE and (S)NRE methods.

For the sequential approach, the simulation budget was split across $5$ rounds. To approximate the epistemic uncertainty we trained $5$ NFs for each simulation budget.

\paragraph{(S)NLE.}
We used \cite{nle1} version of NLE and SNLE algorithm. In line with previous works \citep{lr4, nle1, lueckmann2021benchmarking, npe3}, our neural density estimator is a Masked Autoregressive Flow (MAF) \citep{maf} with 5 autoregressive layers, each has two hidden layers of 50 units each. We used Tanh activation functions. Still in line with previous works, we used Slice Sampling schemes to recover the posterior distribution. Note that this is not the most efficient MCMC to explore high-dimensional or multi-modal spaces. However, since we are in a $6$ almost Gaussian dimensional space this scheme works very well.

\paragraph{(S)NPE.}
We used NPE algorithm as formulated in \cite{npe1} but used as a neural density estimator a MAF instead of a Mixture Density Network (MDN). For SNPE algorithm we use Automatic Posterior Transformation (APT) by \cite{npe3}.  In line with previous works, our neural density estimator is a MAF with 5 autoregressive layers, each has two hidden layers of 50 units each. We used Tanh activation functions. For APT, to compute the atomic proposal, we used $M = 10$ atoms. The computational complexity of APT is $O(M^2)$ and as underlined by \cite{lueckmann2021benchmarking} more atoms are very demanding in terms of memory. In addition, unlike \cite{npe3} we found a difference in training time between $M = 10$ and $M = 100$ atoms. 

Even though APT outperforms previous sequential NPE methods \citep{npe1, npe2}, as reported by the APT paper itself \cite{npe3} and \cite{lr4}, this algorithm can suffer from leakage of posterior mass outside the prior support. To overcome this issue \cite{deistler2022truncated} introduced Truncated Sequential Neural Posterior Estimation (TSNPE).

\paragraph{(S)NRE.}
We used NRE algorithm as in \cite{lr4} and used $K = 10$ class.
In line with previous works \citep{lr4,lueckmann2021benchmarking}, the $K$ multi-class classifier is a residual neural network with two residual blocks of 50 hidden units and ReLU activation functions. Still in line with previous works, we used Slice Sampling schemes to recover the posterior distribution.

\section{Additional convergence plots}

\begin{figure*}[!h]
    \centering
    
    \begin{minipage}{0.45\textwidth}
        \centering
        \includegraphics[width=0.9\linewidth]{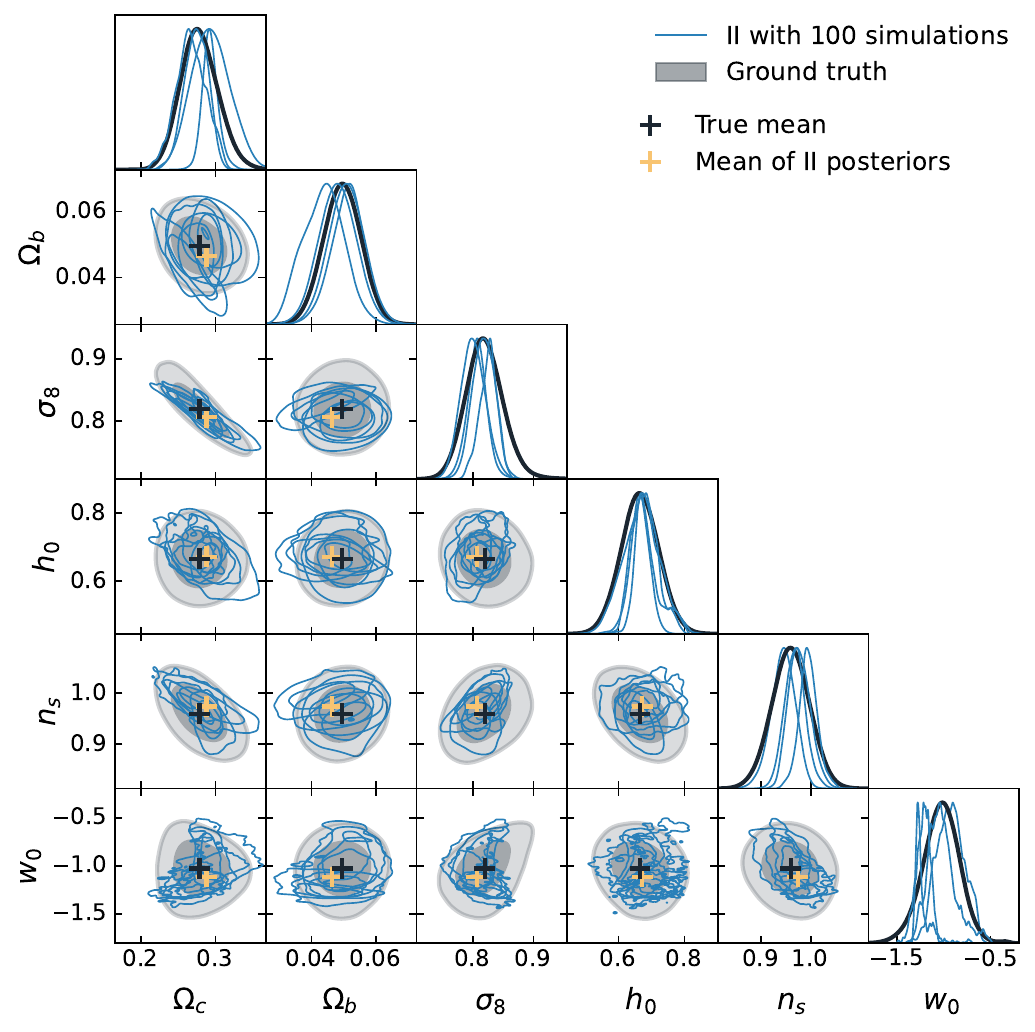}
    \end{minipage}%
    \begin{minipage}{0.45\textwidth}
        \centering
        \includegraphics[width=0.9\linewidth]{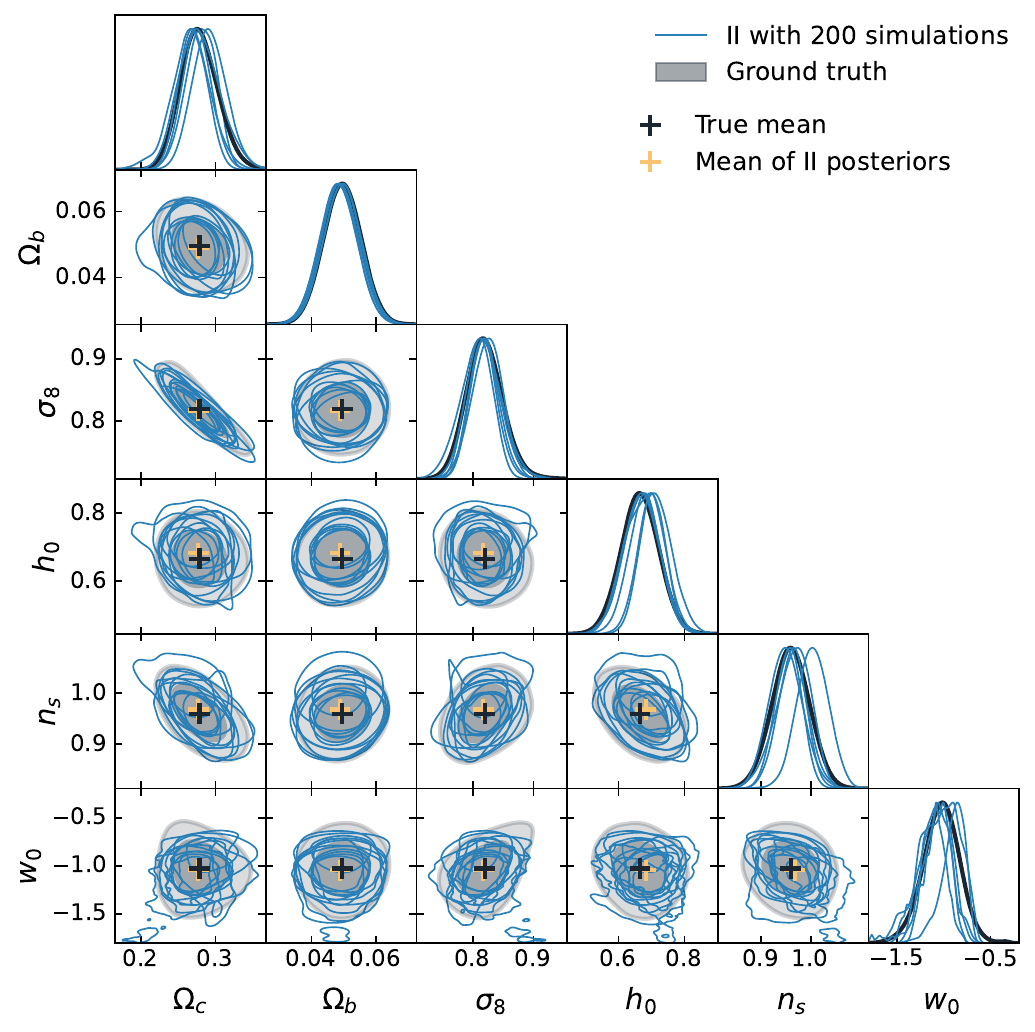}
    \end{minipage}
    
    \medskip
    
    \begin{minipage}{0.45\textwidth}
        \centering
        \includegraphics[width=0.9\linewidth]{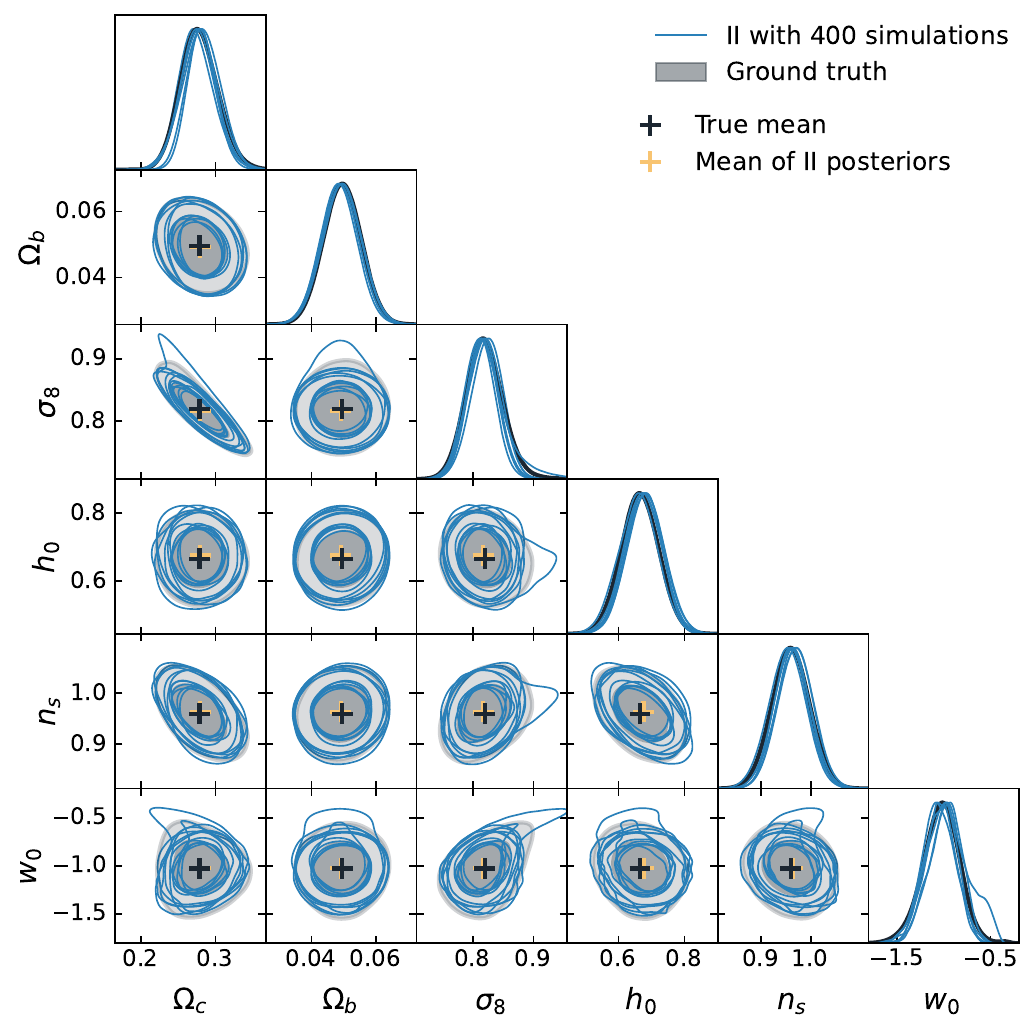}
    \end{minipage}%
    \begin{minipage}{0.45\textwidth}
        \centering
        \includegraphics[width=0.9\linewidth]{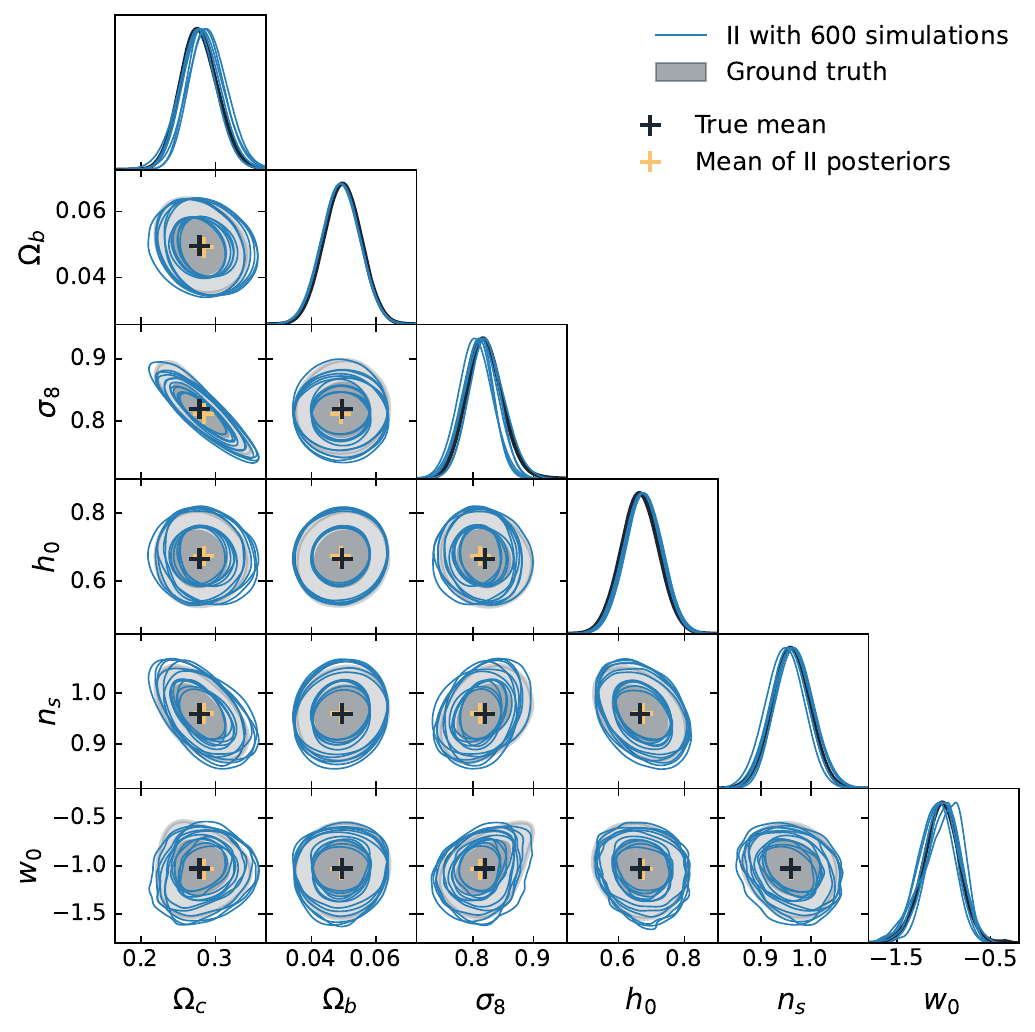}
    \end{minipage}
    
    \medskip
    
    \begin{minipage}{0.45\textwidth}
        \centering
        \includegraphics[width=0.9\linewidth]{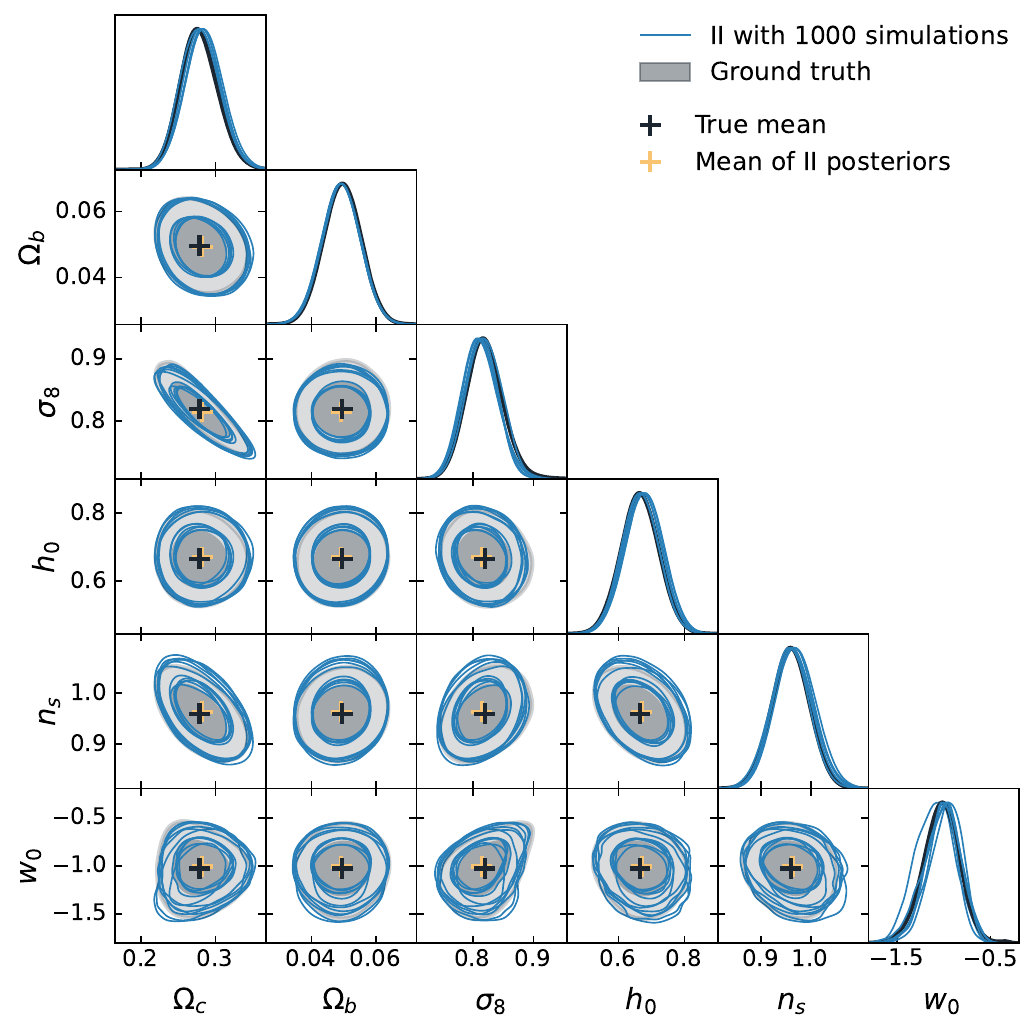}
    \end{minipage}%
    \begin{minipage}{0.45\textwidth}
        \centering
        \includegraphics[width=0.9\linewidth]{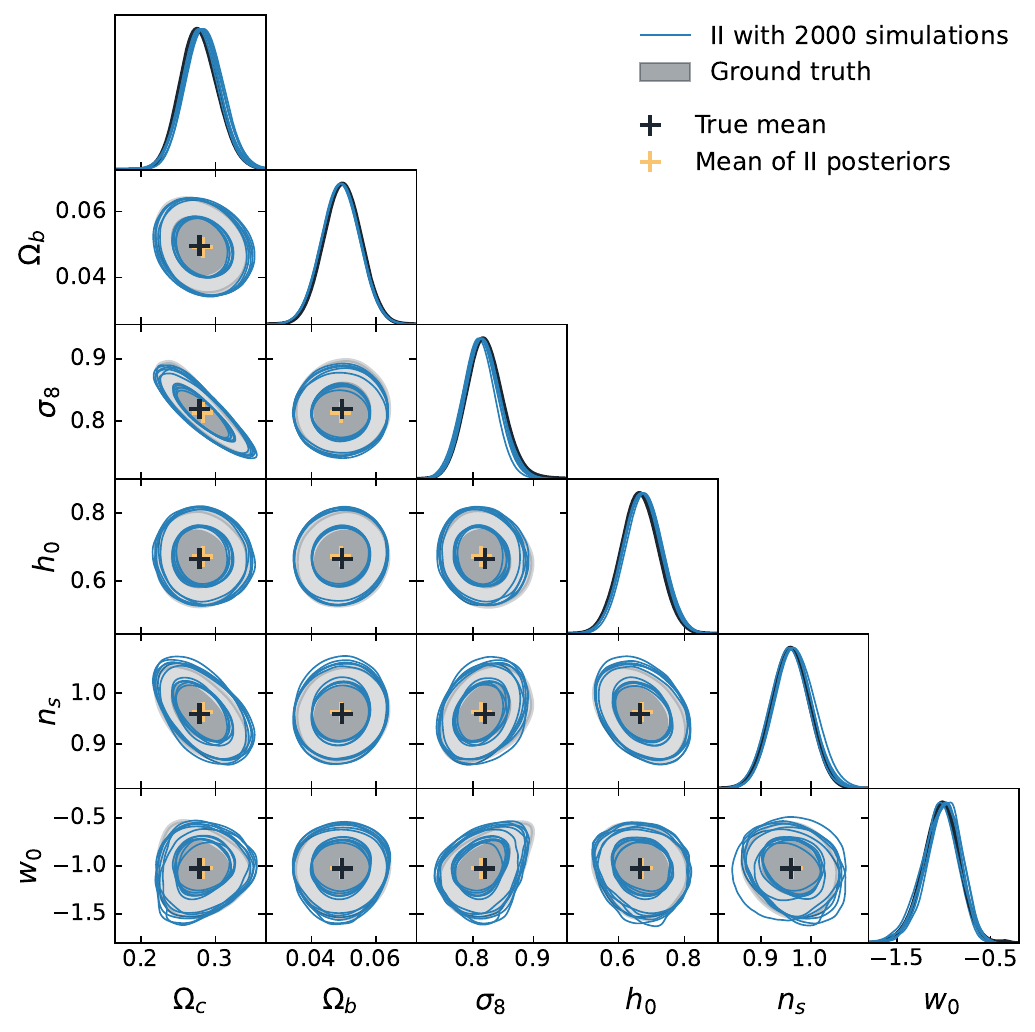}
    \end{minipage}
    
    \caption{{Evolution of implicit inference  posteriors according to the number of simulations used  to train the NF.} The posteriors (blue contours) are approximated using the NLE method with simulations only. We train $5$ NFs with the same architecture where only the initialization of the weights of the NF changes. Each blue contours correspond to the approximated posterior of one of these NFs. 
    The ground truth (black contours) corresponds to the explicit inference posterior of $160\,000$ samples and the black marker corresponds to its mean. The yellow marker corresponds to the mean of the approximated posterior.
    With few simulations (e.g. $100$ simulations) every NF predicts a different posterior and each prediction is overconfident. With a bit more simulations (e.g. $1\,000$ simulations), the posteriors are consistent and similar to the ground truth.}
    \label{fig:nle_contours_evol}
\end{figure*}

\begin{figure*}[h]
        \begin{center}
        \includegraphics[width=\textwidth]{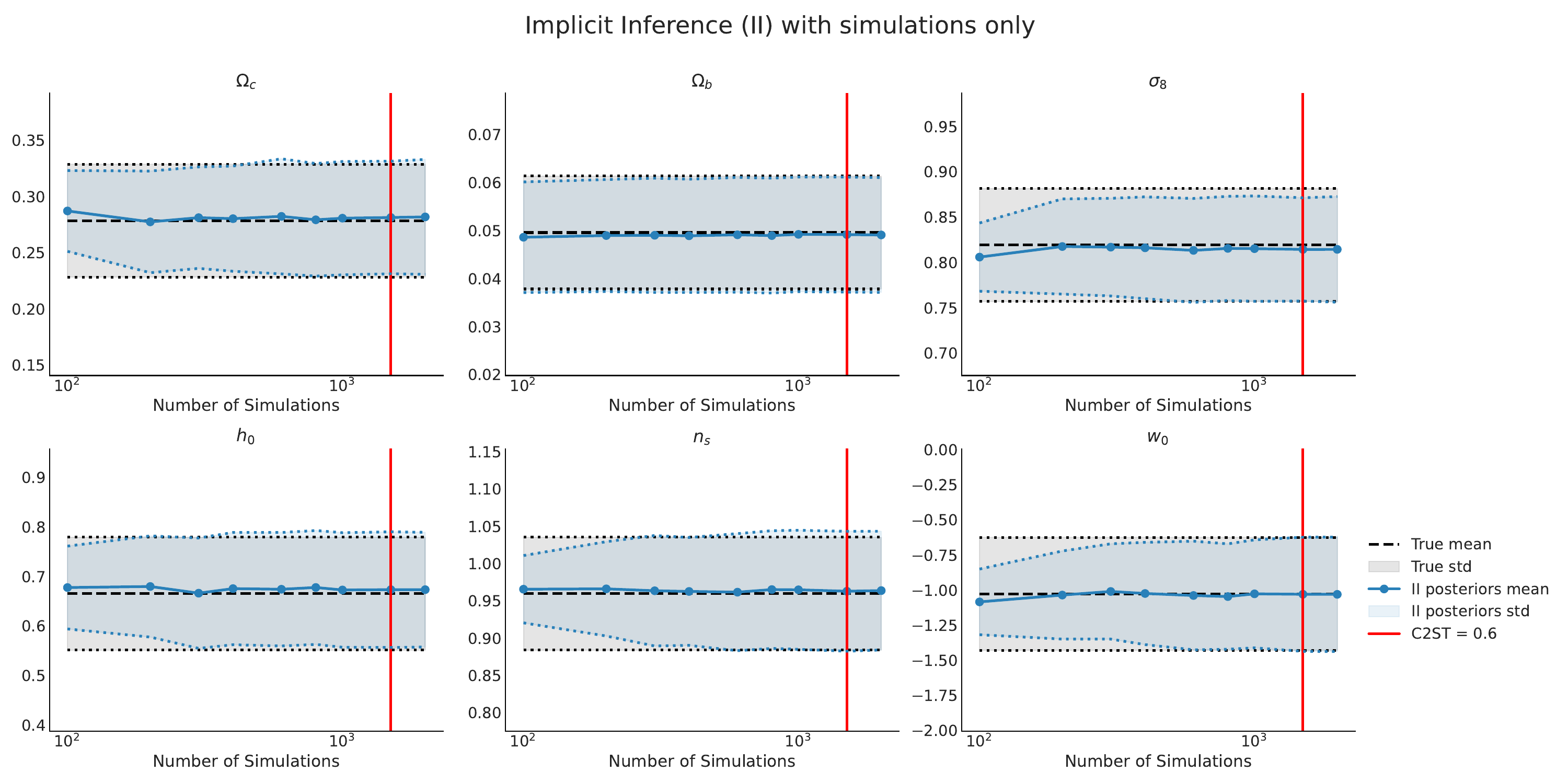}
         \end{center}
         \caption{{Evolution of the mean and standard deviation of the approximated implicit inference posterior as a number of simulations.} The posterior is approximated using the NLE method with only simulations. We train five NFs with the same architecture where only the initialization of the weights of the NF changes. The blue line corresponds to the mean of the five approximated posteriors and the dotted line to the standard deviation. 
        The dashed line corresponds to the mean of the ground truth (the explicit inference posterior of $160\,000$ samples), and the black dotted line to its standard deviation.  The red line corresponds to the number of simulations for which the C2ST is equal to $0.6$ (i.e. assume that the posterior is converged). Note that the C2ST can compare "higher moments" than the first and second moments of two distributions thus these plots cannot serve as direct conclusions.}
        \label{fig:nle_mean_std}
    \end{figure*}

\begin{figure*}[h]
        \begin{center}
        \includegraphics[width=\textwidth]{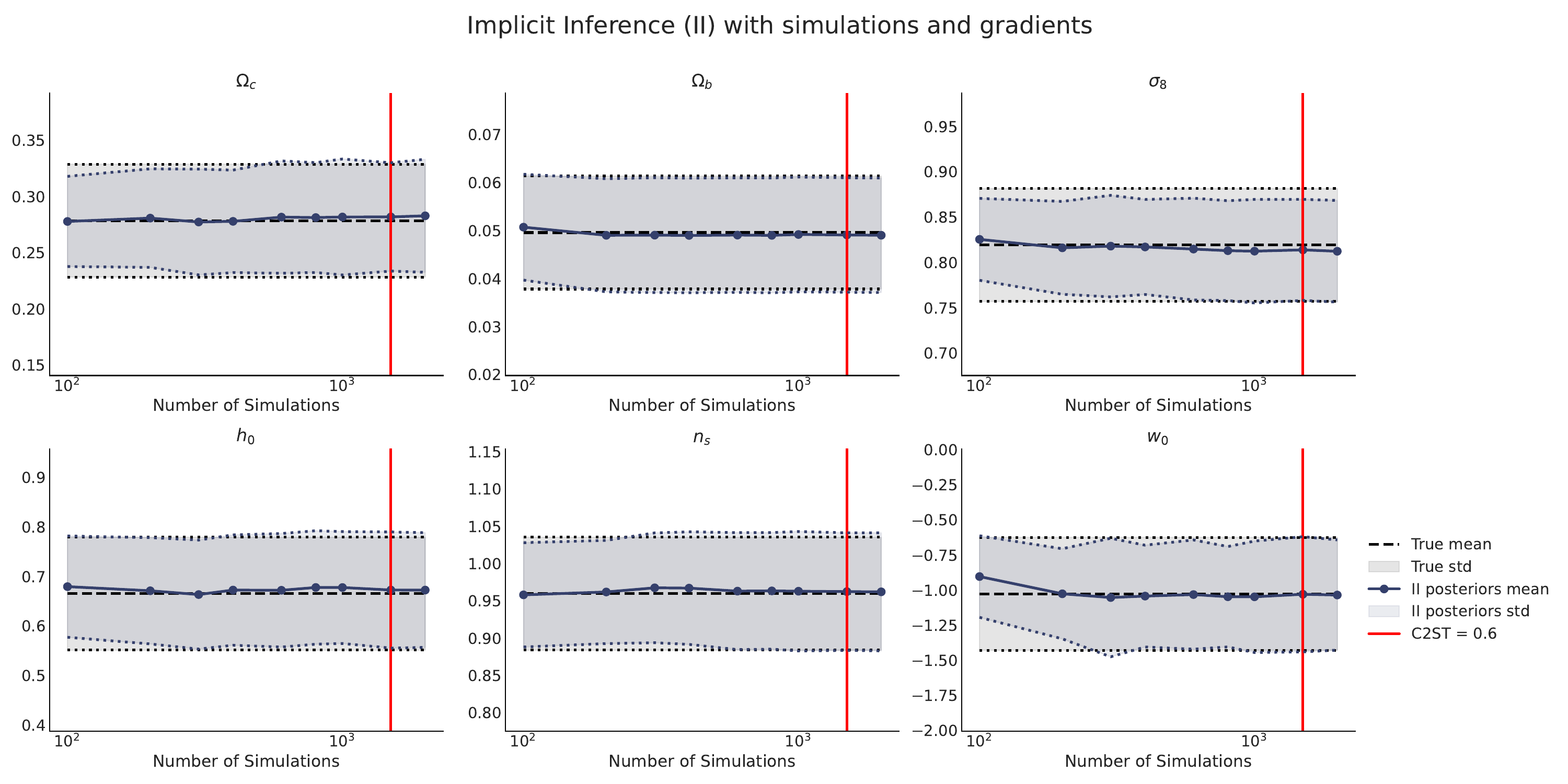}
        \caption{Same as the previous figure but this time the posterior is approximated using the $\partial$NLE method with simulations and gradients.}
        \label{fig:partialnle_mean_std}
         \end{center}
    \end{figure*}

\begin{figure*}[h]
        \begin{center}
        \includegraphics[width=\textwidth]{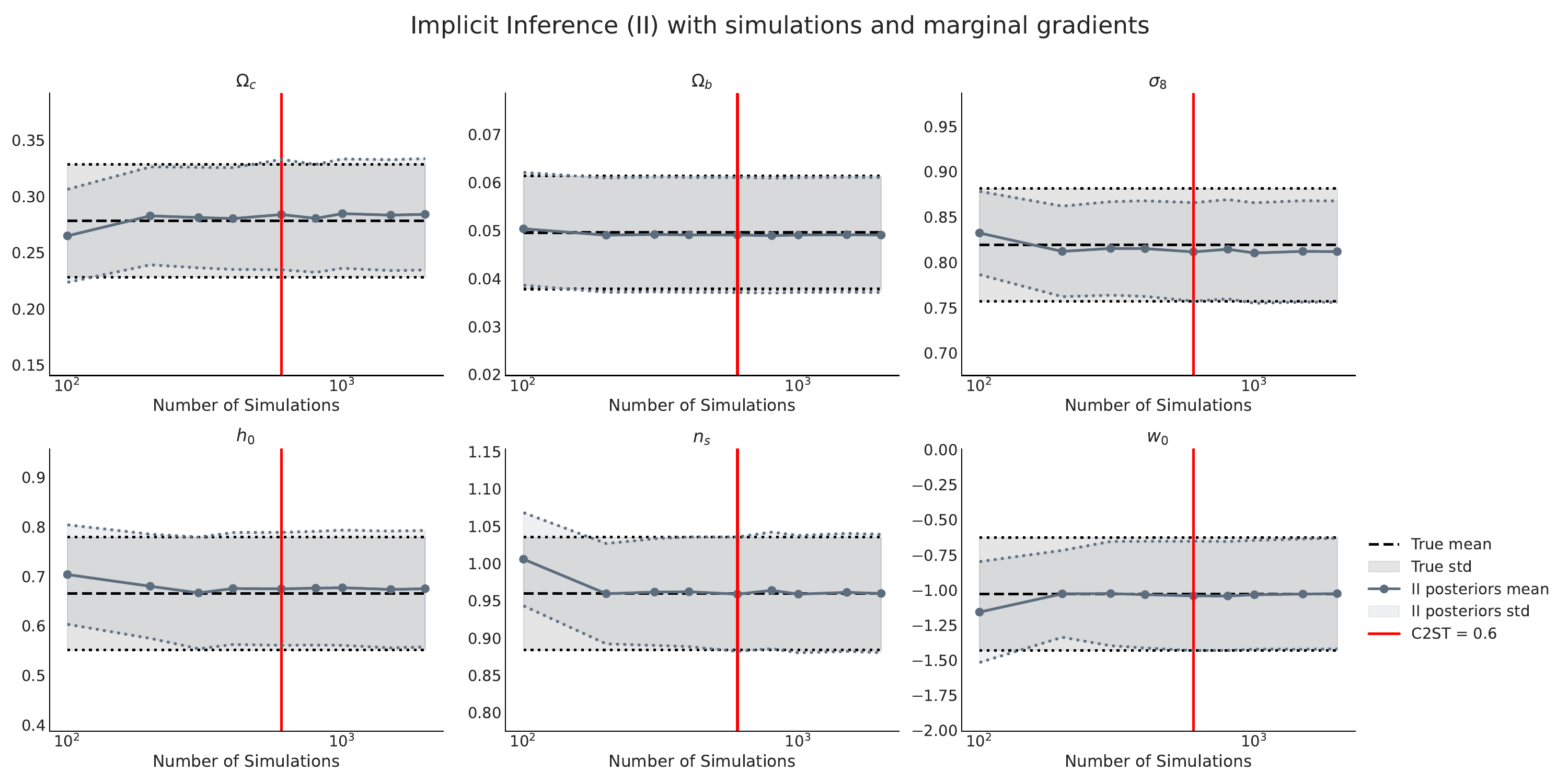}
        \caption{Same as the previous figure but this time the posterior is approximated using the $\partial$NLE method with simulations and marginal gradients.}
        \label{fig:marginalpartialnle_mean_std}
         \end{center}
    \end{figure*}

\begin{figure*}[h]
        \begin{center}
        \includegraphics[width=\textwidth]{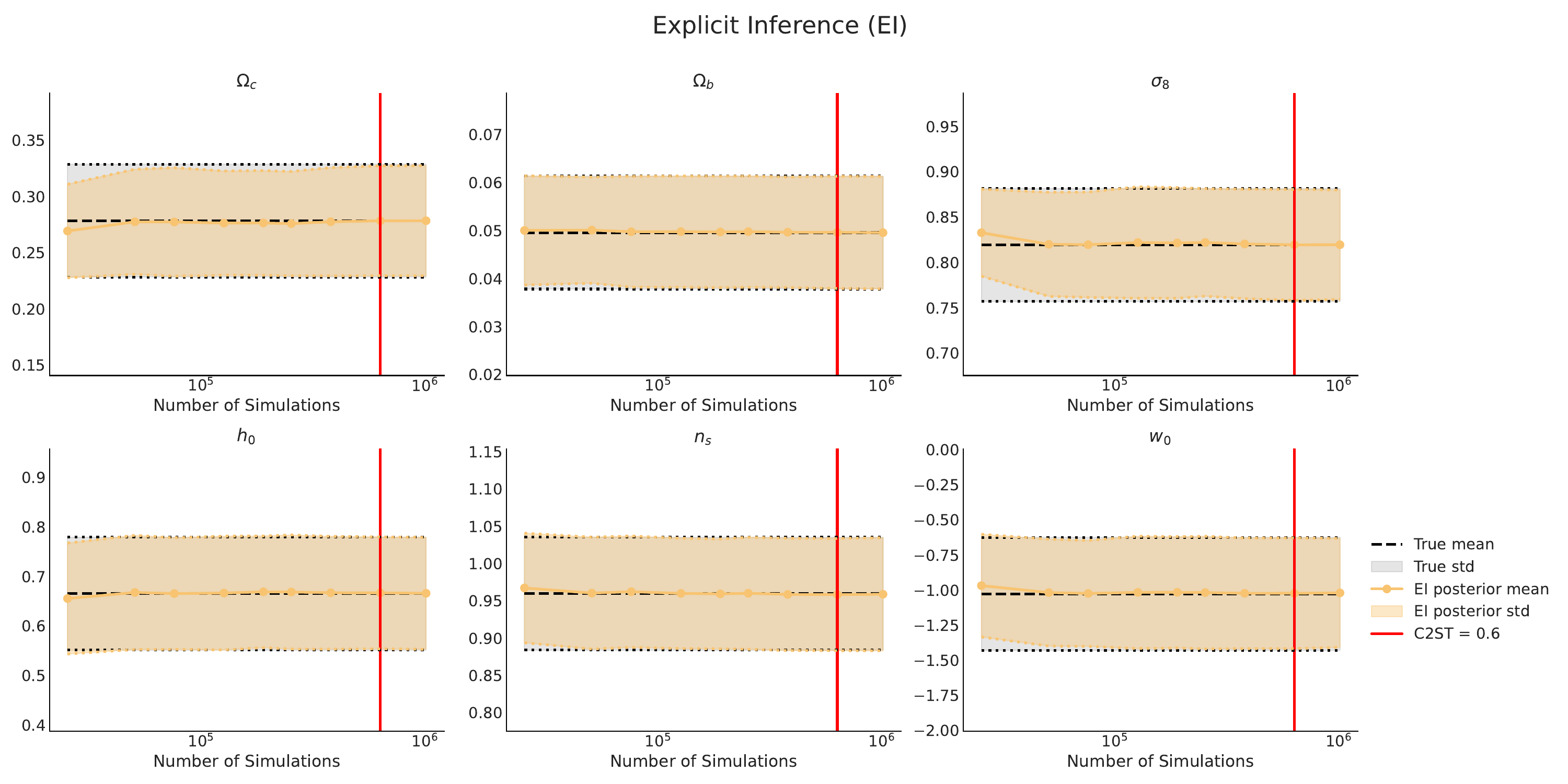}
        \caption{{Evolution of the mean and standard deviation of the explicit inference posterior as a number of simulations.} To get our $160\,000$ posterior samples (our ground truth) we use the NUTS algorithm. For each simulation budget $N$, we select the first $N$ samples of this ground truth and compute its mean and standard deviation. The yellow line and dotted line correspond respectively to the mean and standard deviation.
        The black dashed line corresponds to the mean of the ground truth (of $160\,000$ posterior samples), and the black dotted line to its standard deviation.  The red line corresponds to the number of simulations for which the C2ST is equal to $0.6$ (i.e. assume that the posterior is converged). Note that the C2ST can compare "higher moments" than the first and second moments of two distributions thus these plots cannot serve as direct conclusions.}
        \label{fig:ei_mean_std}
         \end{center}
    \end{figure*}

\begin{figure*}[!h]
    \centering
    
    \begin{minipage}{0.45\textwidth}
        \centering
        \includegraphics[width=0.9\linewidth]{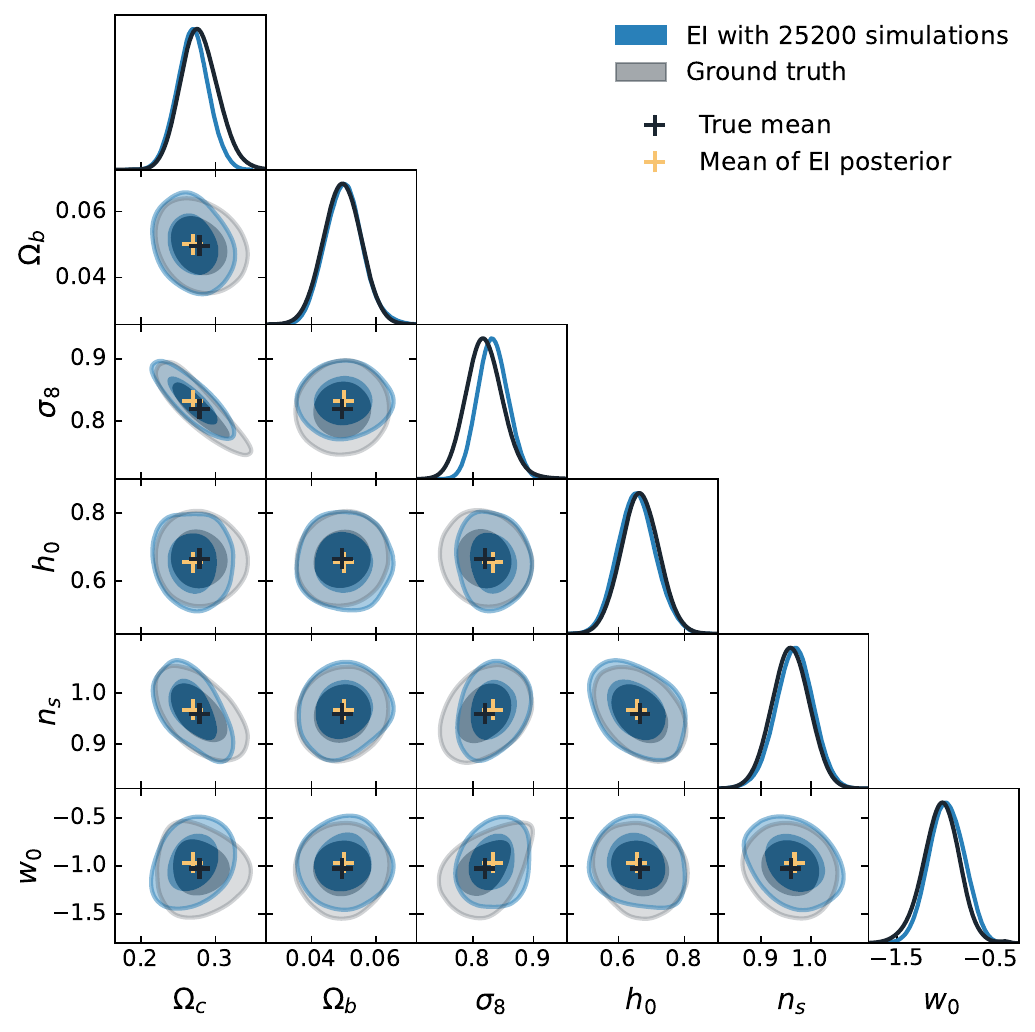}
        \label{fig:plot1}
    \end{minipage}%
    \begin{minipage}{0.45\textwidth}
        \centering
        \includegraphics[width=0.9\linewidth]{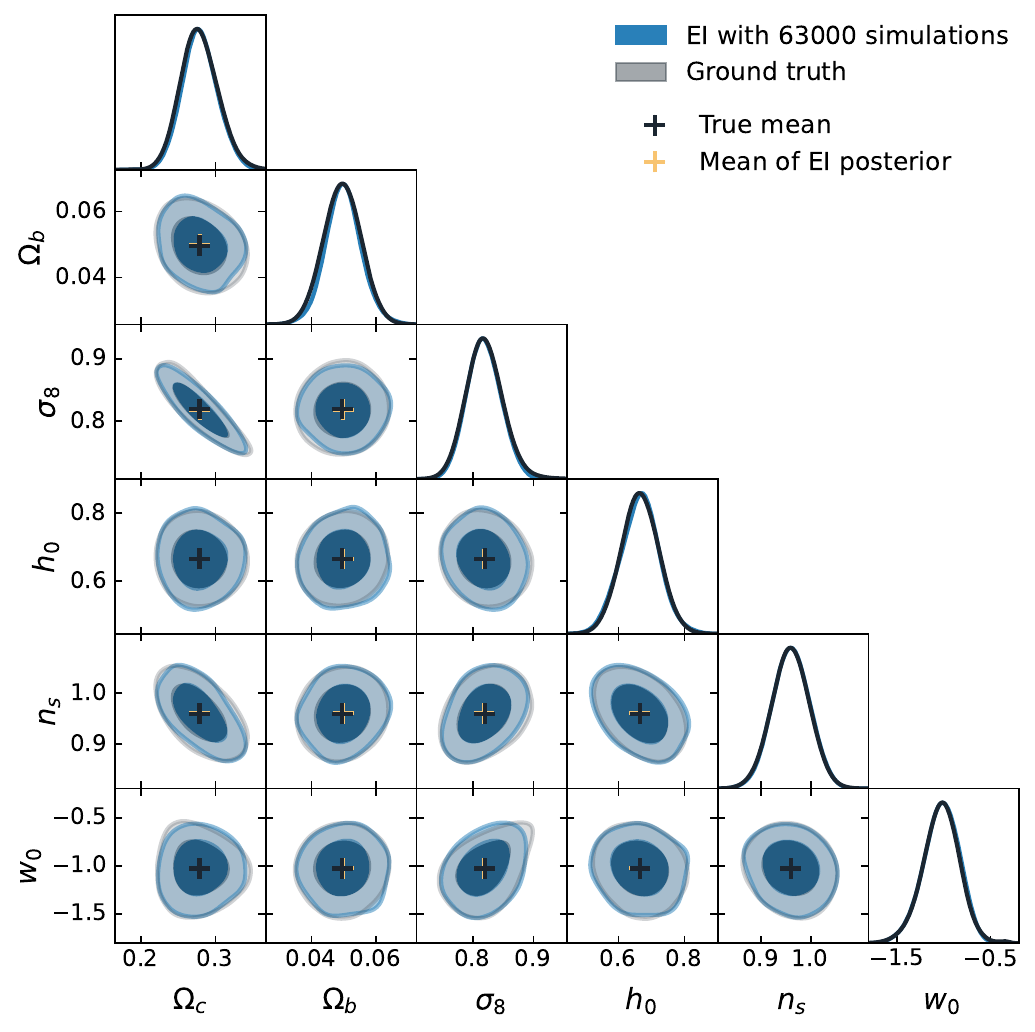}
        \label{fig:plot2}
    \end{minipage}
    
    \medskip
    
    \begin{minipage}{0.45\textwidth}
        \centering
        \includegraphics[width=0.9\linewidth]{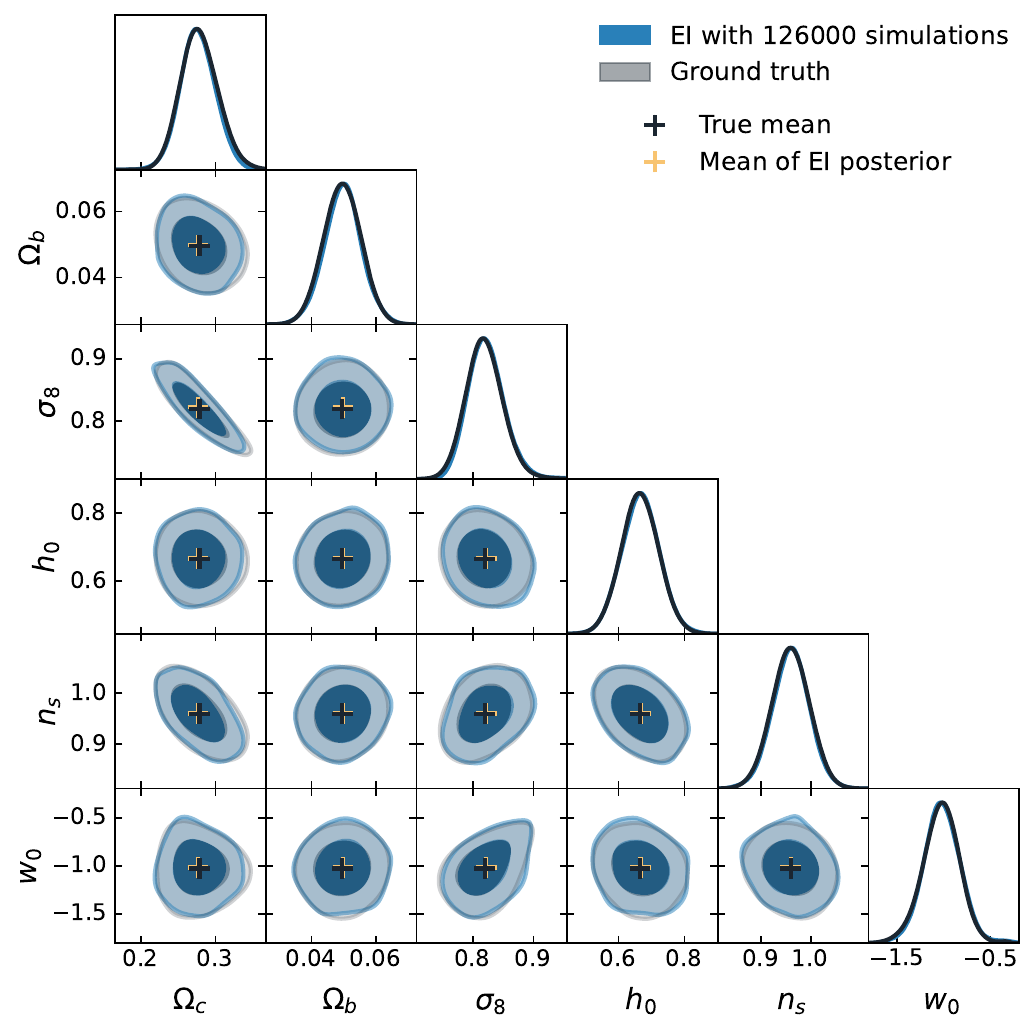}
        \label{fig:plot3}
    \end{minipage}%
    \begin{minipage}{0.45\textwidth}
        \centering
        \includegraphics[width=0.9\linewidth]{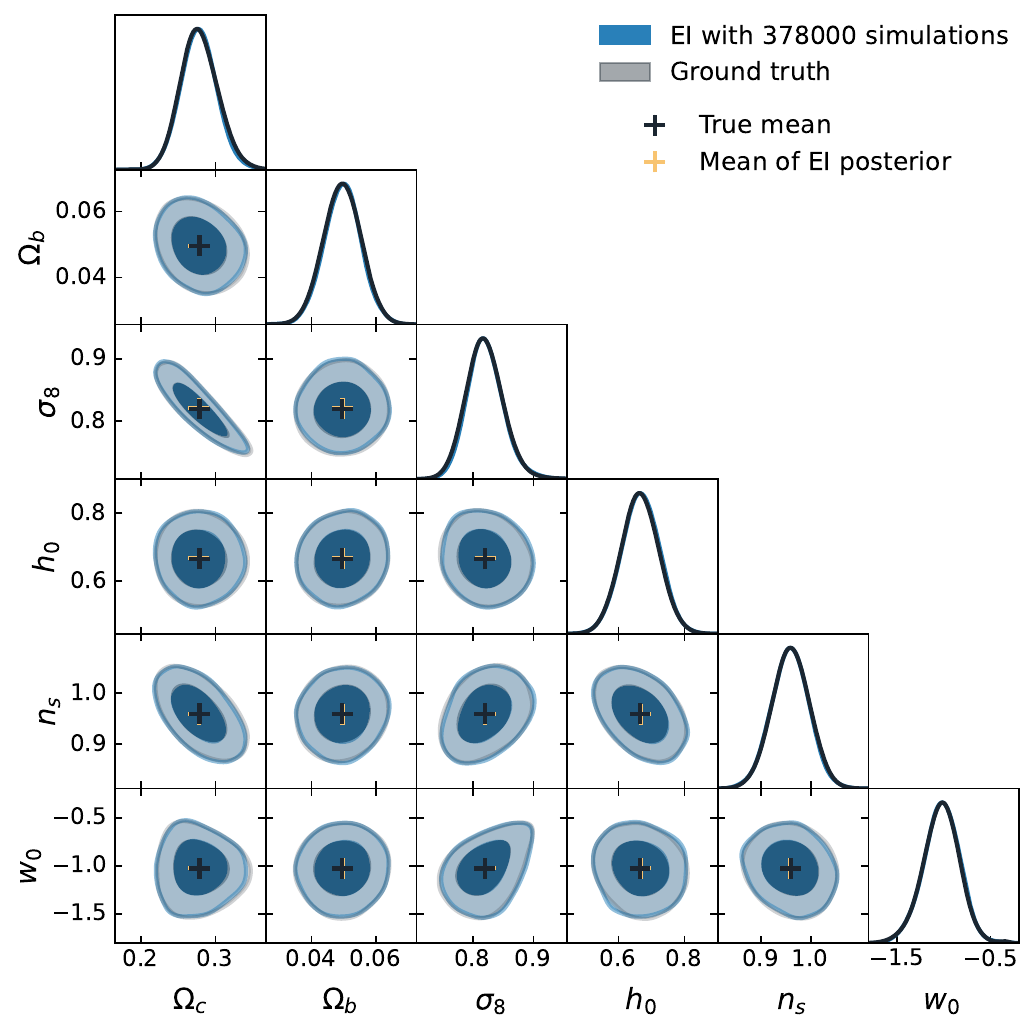}
        \label{fig:plot4}
    \end{minipage}
    
    \medskip
    
    \begin{minipage}{0.45\textwidth}
        \centering
        \includegraphics[width=0.9\linewidth]{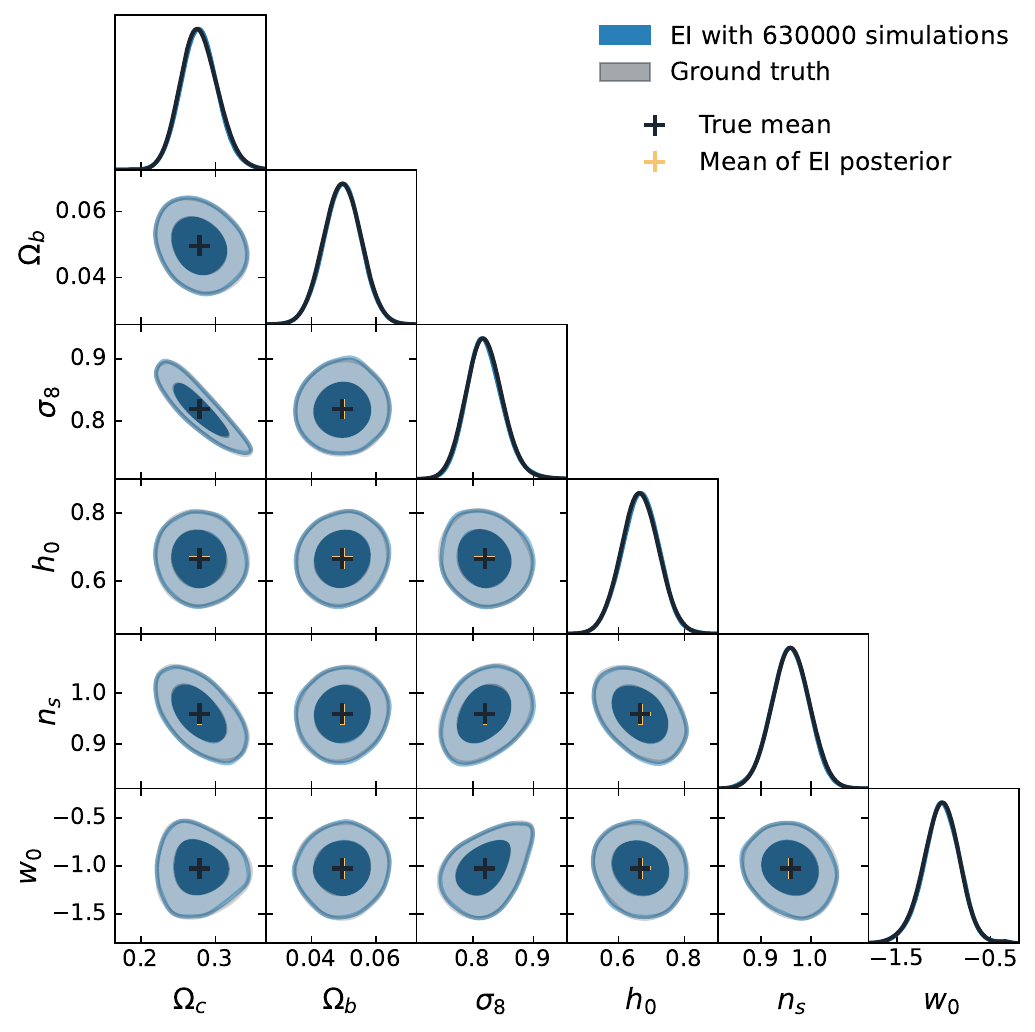}
        \label{fig:plot5}
    \end{minipage}%
    \begin{minipage}{0.45\textwidth}
        \centering
        \includegraphics[width=0.9\linewidth]{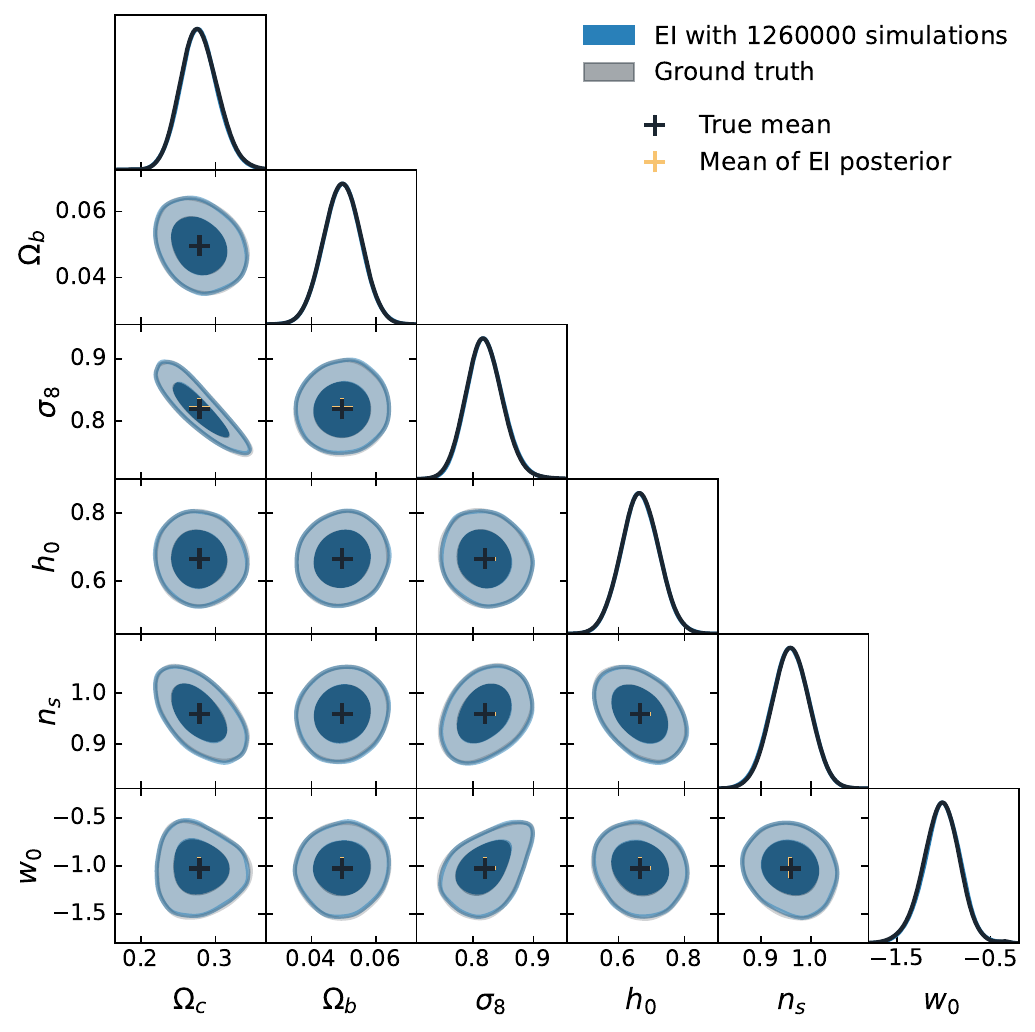}
        \label{fig:plot6}
    \end{minipage}
    
    \caption{{Evolution in terms of simulations of the explicit inference approximated posterior.} The ground truth (black contours) corresponds to the explicit inference posterior of $160\,000$ samples. The blue contours denote the first $N$ samples of the ground truth. Note that to get one sample our NUTS algorithm requires $126$ simulations.
    The black marker corresponds to the mean of the ground truth. The yellow one, to the mean of the approximated posterior.}
    \label{fig:ei_contour_plots}
\end{figure*}

\begin{figure*}[!h]
    \centering
    \begin{minipage}{0.43\textwidth}
        \centering
        \includegraphics[width=0.9\linewidth]{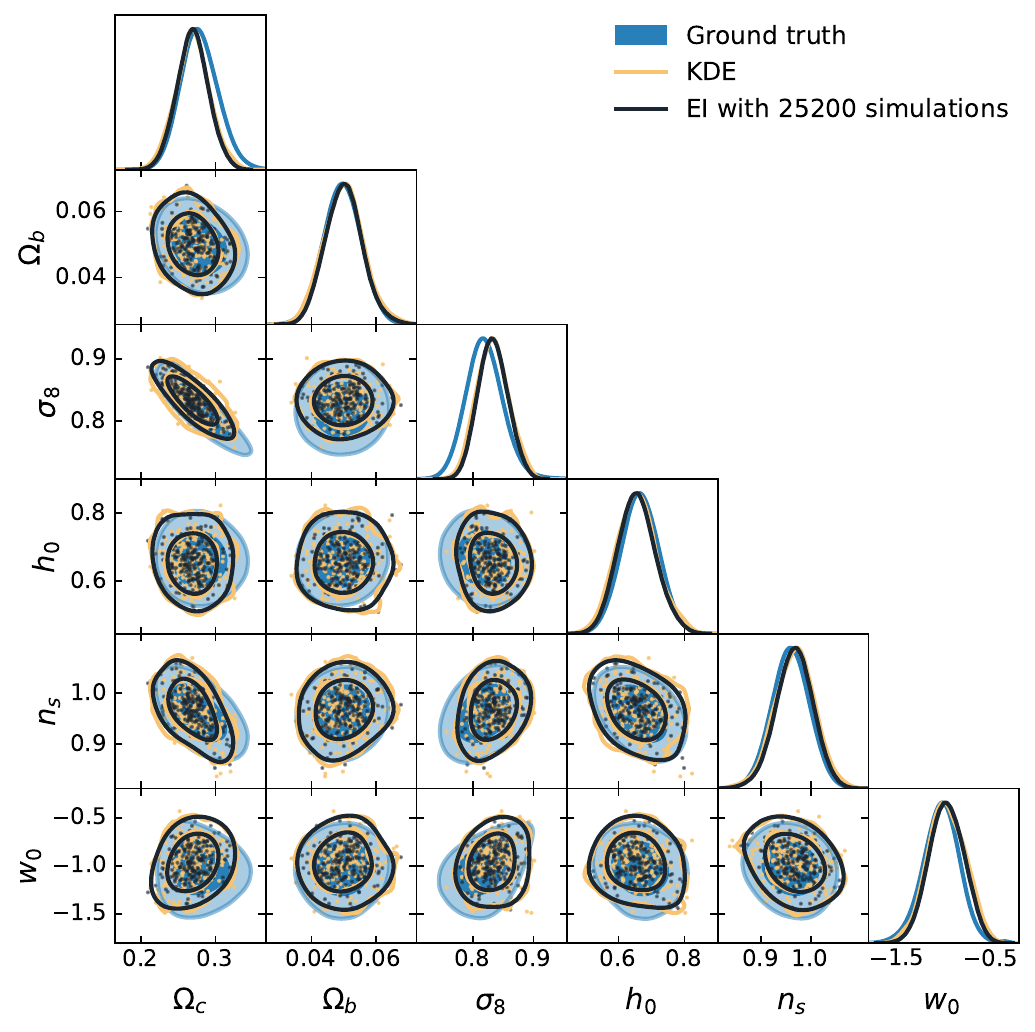}
        \label{fig:plot1}
    \end{minipage}%
    \begin{minipage}{0.43\textwidth}
        \centering
    \includegraphics[width=0.9\linewidth]{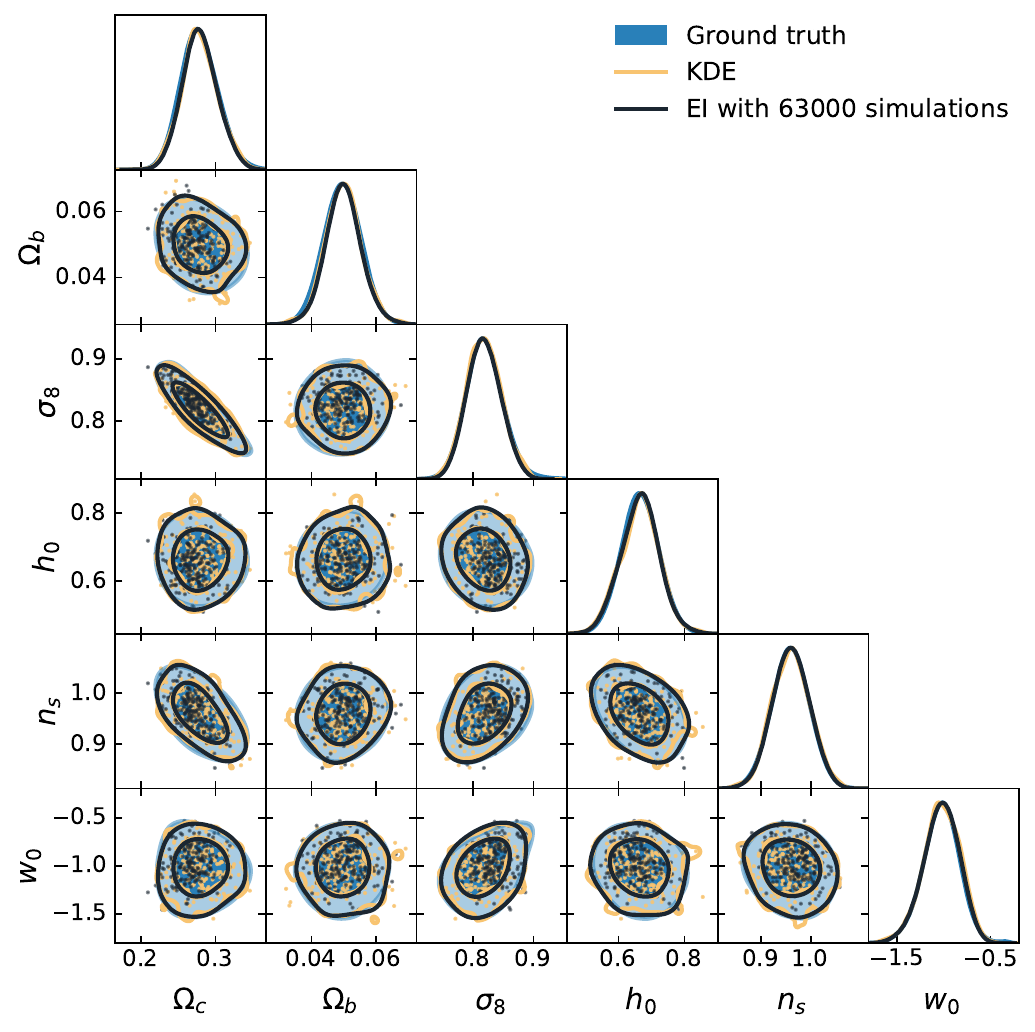}
        \label{fig:plot2}
    \end{minipage}
    
    \medskip
    
    \begin{minipage}{0.43\textwidth}
        \centering
        \includegraphics[width=0.9\linewidth]{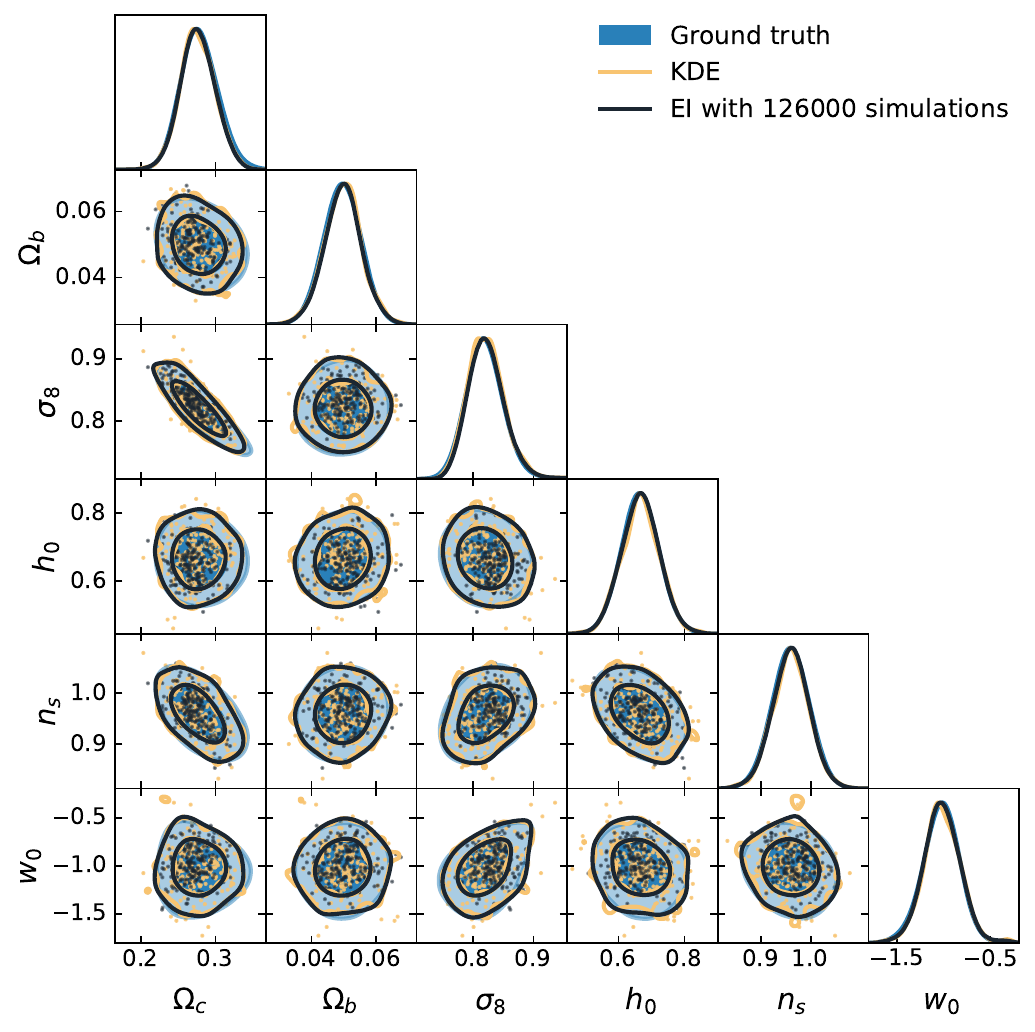}
        \label{fig:plot1}
    \end{minipage}%
    \begin{minipage}{0.43\textwidth}
        \centering
    \includegraphics[width=0.9\linewidth]{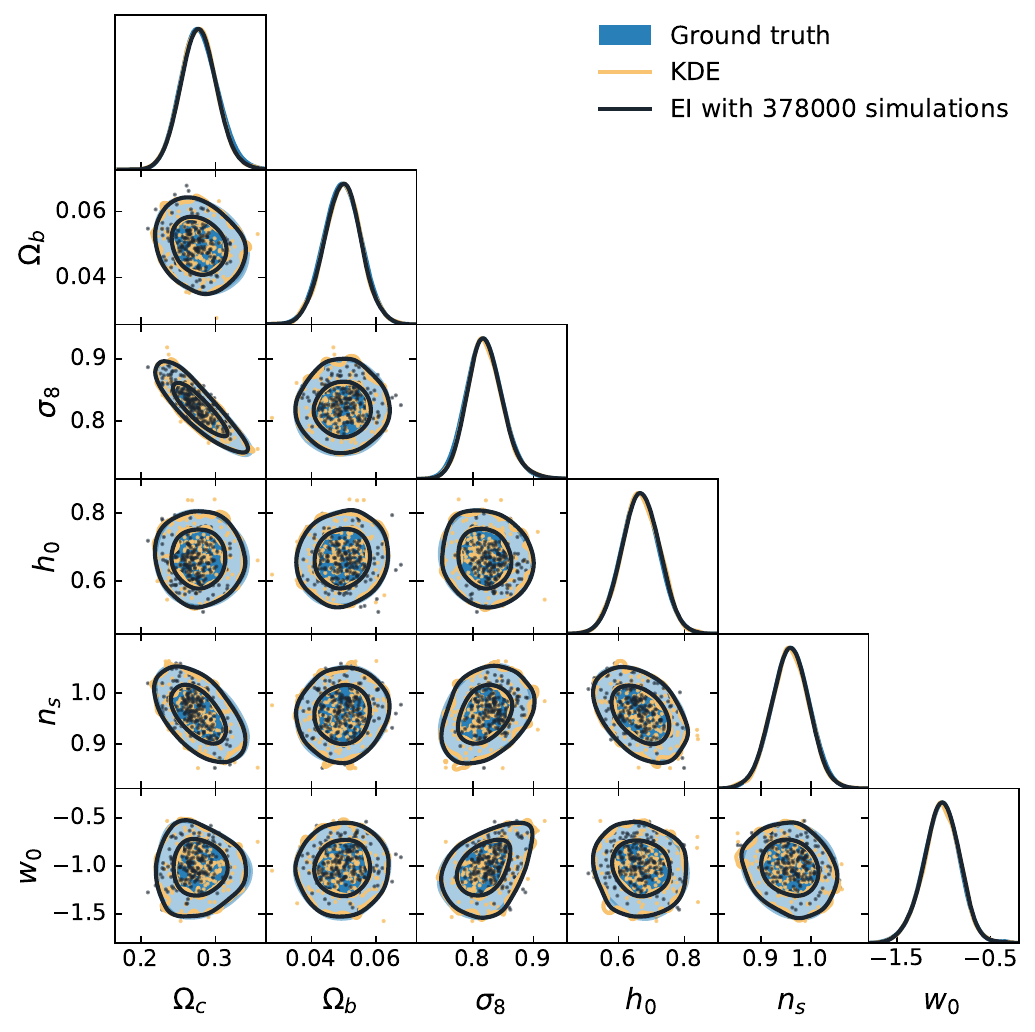}
        \label{fig:plot2}
    \end{minipage}
    
    \medskip
    
    \begin{minipage}{0.43\textwidth}
        \centering
        \includegraphics[width=0.9\linewidth]{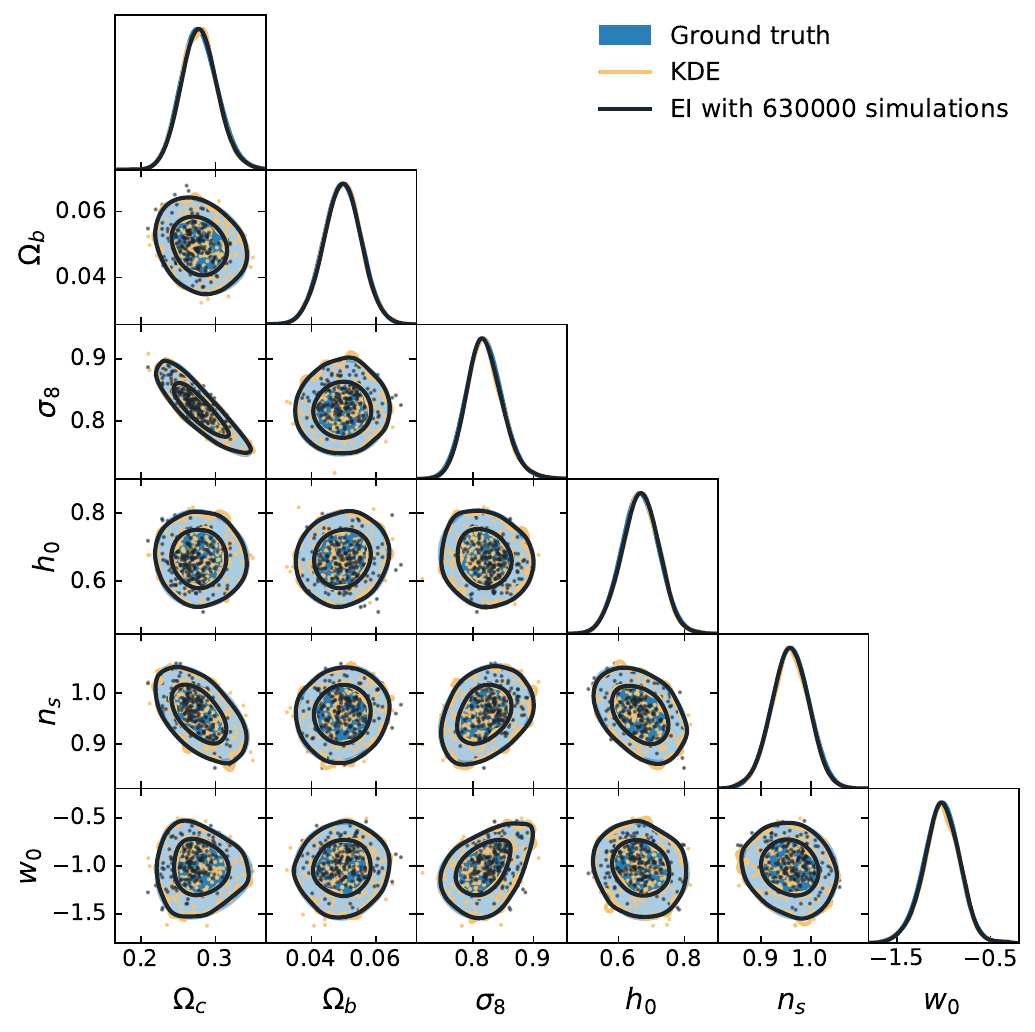}
        \label{fig:plot3}
    \end{minipage}%
    \begin{minipage}{0.43\textwidth}
        \centering
\includegraphics[width=0.9\linewidth]{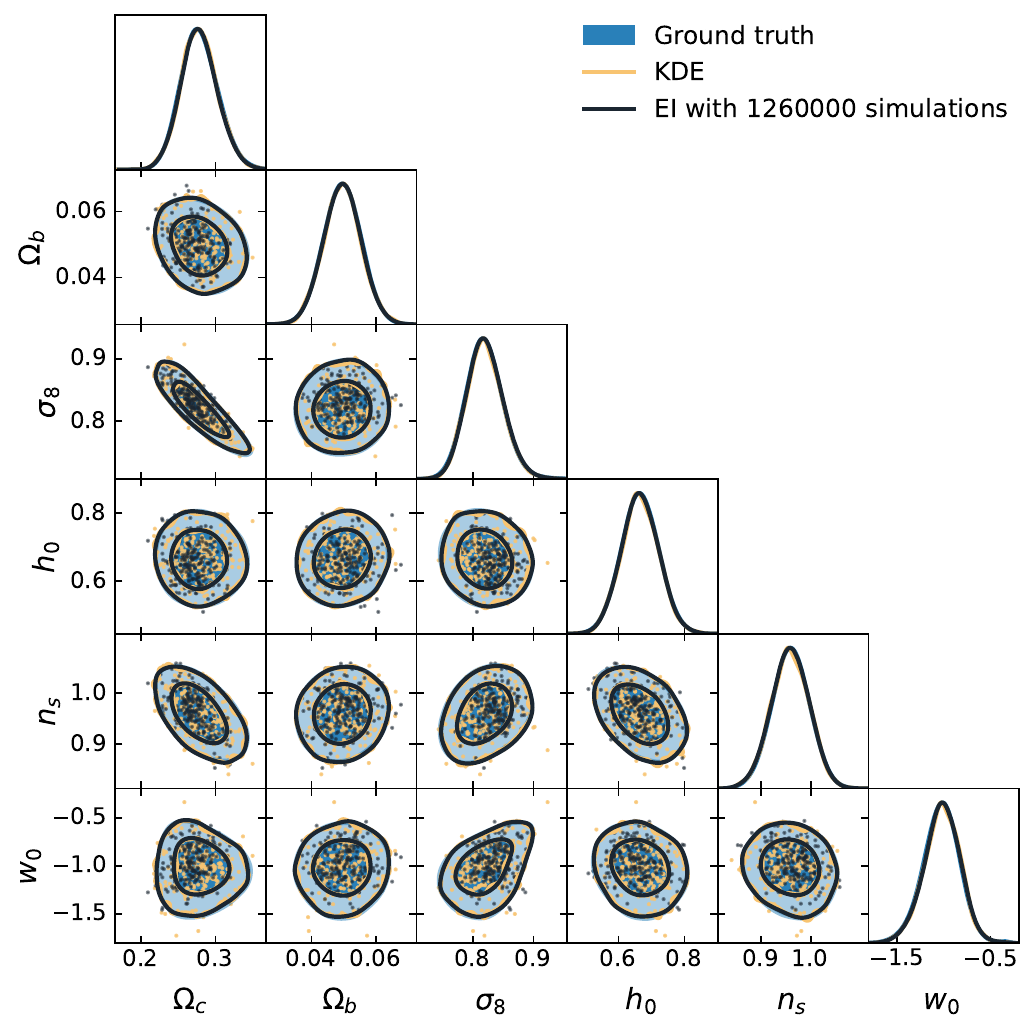}
        \label{fig:plot4}
    \end{minipage}
    \caption{{Samples of the KDE used to compute C2ST metric.} In this paper we use the C2ST metric to evaluate the convergence of inference methods. This metric requires an equal number of simulations of the two distributions to be compared. To use the C2ST metric to benchmark the explicit inference method, inspired by the construction of contour plots that smooth the distribution (such as the one proposed by GetDist), we use a KDE to generate new samples. Specifically, we use a Gaussian kernel and set the bandwidth to match what GetDist would display for every number of explicit inference posterior samples (black contours). The yellow contours correspond to the contours obtained when fitting the $N$ samples of explicit inference posterior and generating $20\,000$ samples from the KDE. We use a very small smoothing scaling value to display the posterior contours of the KDE with GetDist. The blue contours denote the ground truth of $160\,000$ samples. }  
    \label{fig:kde_proof}
\end{figure*}

We provide additional results showing the convergence of inference methods. Figure \ref{fig:nle_contours_evol} shows the contours evolution of the implicit inference posteriors approximated with NLE method. 
Figures \ref{fig:nle_mean_std}, \ref{fig:marginalpartialnle_mean_std}, \ref{fig:partialnle_mean_std}, and \ref{fig:ei_mean_std} show the evolution of the approximated mean and standard deviation of the posteriors approximated with NLE, $\partial$NLE with joint gradients and marginal gradient, and the explicit inference methods. Figure \ref{fig:ei_contour_plots} shows the contours evolution of the explicit inference posterior. Finally, Fig. \ref{fig:kde_proof} displays the KDE approximation used to compute the C2ST metric of explicit inference method.


\end{appendix}

\end{document}